\newif\ifistechreport
\istechreportfalse

\newif\ifisanon
\isanonfalse

\ifisanon
  \documentclass[10pt,sigconf,letterpaper,nonacm,anonymous]{acmart}
\else
  \documentclass[10pt,sigconf,letterpaper,nonacm]{acmart}
\fi

\usepackage[utf8]{inputenc}

\usepackage{booktabs} %
\usepackage[inline]{enumitem} %
\usepackage{caption}
\usepackage[multidot]{grffile}
\usepackage{graphics}
\usepackage{tikz-imagelabels}
\imagelabelset {
  border thickness = 0.4pt,
  arrow thickness = 0.3pt,
  arrow distance = 0pt,
  tip size = 0.4pt,
  outer dist = 2mm,
}
\usepackage{float}
\usepackage{subfig}
\usepackage{hyperref}
\usepackage{amsmath}

\usepackage{xspace}
\usepackage{algpseudocode}
\usepackage{tikz}
\usepackage{multirow}
\usepackage{mathtools}
\usepackage{colortbl}

\usepackage{graphicx}
\usepackage{xcolor}

\usepackage[export]{adjustbox}

\usepackage{algorithm}

\algnewcommand{\LineComment}[1]{\State \(\triangleright\) #1}

\newcommand{\DGreenCell}{\cellcolor[HTML]{99ee77}}
\newcommand{\LGreenCell}{\cellcolor[HTML]{ccee99}}
\newcommand{\YellowCell}{\cellcolor[HTML]{fff9c4}}
\newcommand{\LRedCell}{\cellcolor[HTML]{f0abab}}

\usepackage{xurl}

\makeatletter
  \renewcommand\paragraph{\@startsection{paragraph}{4}{\parindent}%
    {-.1\baselineskip \@plus -2\p@ \@minus -.2\p@}%
    {-3.5\p@}%
    {\@parfont\@adddotafter}}
\makeatother

\ifisanon
  \newcommand{\groot}{Y-Root\xspace}
  \newcommand{\broot}{X-Root\xspace}

  \newcommand{\EnterpriseUniversity}{Our University\xspace}
  \newcommand{\EnterpriseU}{Our Univ.\xspace}
  \newcommand{\EnterpriseCenic}{Academic Regional Network A\xspace} 
  \newcommand{\EnterpriseLN}{Academic Regional Network B\xspace} 
  \newcommand{\EnterpriseInternetTwo}{Academic National Network C\xspace} 
  \newcommand{\EnterpriseHE}{Hurricane Electric\xspace} 
  \newcommand{\EnterpriseNTT}{NTT\xspace} 
\else
  \newcommand{\groot}{G-Root\xspace}
  \newcommand{\broot}{B-Root\xspace}

  \newcommand{\EnterpriseUniversity}{the University of Southern California\xspace}
  \newcommand{\EnterpriseU}{USC\xspace}
  \newcommand{\EnterpriseCenic}{CENIC (a regional academic network)\xspace} 
  \newcommand{\EnterpriseInternetTwo}{Internet2\xspace} 
  \newcommand{\EnterpriseLN}{LosNettos(a regional network)\xspace} 
  \newcommand{\EnterpriseHE}{Hurricane Electric\xspace} 
  \newcommand{\EnterpriseNTT}{NTT\xspace} 
\fi

\definecolor{Set3c0}{rgb}{0.55294117647058827,0.82745098039215681,0.7803921568627451}
\colorlet{Set3c0p}{Set3c0!50!white}
\definecolor{Set3c1}{rgb}{1.0,1.0,0.70196078431372544}
\colorlet{Set3c1p}{Set3c1!50!white}
\definecolor{Set3c2}{rgb}{0.74509803921568629,0.72941176470588232,0.85490196078431369}
\colorlet{Set3c2p}{Set3c2!50!white}
\definecolor{Set3c3}{rgb}{0.98431372549019602,0.50196078431372548,0.44705882352941179}
\colorlet{Set3c3p}{Set3c3!50!white}
\definecolor{Set3c4}{rgb}{0.50196078431372548,0.69411764705882351,0.82745098039215681}
\colorlet{Set3c4p}{Set3c4!50!white}
\definecolor{Set3c5}{rgb}{0.99215686274509807,0.70588235294117652,0.3843137254901961 }
\colorlet{Set3c5p}{Set3c5!50!white}
\definecolor{Set3c6}{rgb}{0.70196078431372544,0.87058823529411766,0.41176470588235292}
\colorlet{Set3c6p}{Set3c6!50!white}
\definecolor{Set3c7}{rgb}{0.9882352941176471,0.80392156862745101,0.89803921568627454}
\colorlet{Set3c7p}{Set3c7!50!white}
\definecolor{Set3c8}{rgb}{0.85098039215686272,0.85098039215686272,0.85098039215686272}
\colorlet{Set3c8p}{Set3c8!50!white}
\definecolor{Set3c9}{rgb}{0.73725490196078436,0.50196078431372548,0.74117647058823533}
\colorlet{Set3c9p}{Set3c9!50!white}
\definecolor{Set3c10}{rgb}{0.8,0.92156862745098034,0.77254901960784317}
\colorlet{Set3c10p}{Set3c10!50!white}
\definecolor{Set3c11}{rgb}{1.0,0.92941176470588238,0.43529411764705883}
\colorlet{Set3c11p}{Set3c11!50!white}
\newcommand{\SCLstack}{\colorbox{Set3c0p}{SCL}\xspace}
\newcommand{\ARIstack}{\colorbox{Set3c1p}{ARI}\xspace}
\newcommand{\AMSstack}{\colorbox{Set3c4p}{AMS}\xspace}
\newcommand{\IADstack}{\colorbox{Set3c3p}{IAD}\xspace}
\newcommand{\SINstack}{\colorbox{Set3c5p}{SIN}\xspace}
\newcommand{\LAXstack}{\colorbox{Set3c6p}{LAX}\xspace}

\newcommand{\PhiModes}[2]{\Phi(M_{\Vs{#1}},M_{\Vs{#2}})}

  \setcopyright{rightsretained}
\ifistechreport
  \copyrightyear{2024}
  \acmYear{2024}
  \acmDOI{}
  \acmConference{arXiv}{June~2024}{Marina del Rey, California, USA}
  \acmISBN{}
\else
  \copyrightyear{2024}
  \acmYear{2024}
  \acmDOI{}
  \acmConference{Under Review (submission)}{May.~2025}{Wisconsin, USA}
  \acmISBN{}
\fi

\makeatletter
\newcommand\EatSpacesHack{\@bsphack\@esphack}
\makeatother

\iftrue
  \renewcommand\comment[1]{\textbf{\sffamily\color{blue}[xxx: #1]}}
  \newcommand\sx[1]{\textbf{\sffamily\color{purple}[xxx: #1]}}
  \newcommand\PostSubmission[1]{\textbf{\sffamily\color{red}[xxx: #1]}}
  \newcommand\reviewfix[1]{{\sffamily\color{green}[RF: #1]}}
\else
  \renewcommand\comment[1]{\EatSpacesHack}
  \newcommand\sx[1]{\EatSpacesHack}
  \newcommand\PostSubmission[1]{\EatSpacesHack}
  \newcommand\reviewfix[1]{\EatSpacesHack}
\fi

\newcommand{\V}[1]{\mbox{\textit{#1}}}  %
\newcommand{\Vs}[1]{\mbox{\textit{\footnotesize #1}}} %
\newcommand{\Vr}[1]{\mbox{#1}}  %

\def\Snospace~{\S{}}

\usepackage{amsmath, amsfonts}
\usepackage{tikz}
\usetikzlibrary{positioning, arrows.meta}

\hypersetup{
	colorlinks,%
	citecolor=[rgb]{0.7,0,0},
	filecolor=[rgb]{0.2,0.2,0.2},
	linkcolor=[rgb]{0,0,0.7},
	urlcolor=[rgb]{0.7,0,0},
}

\definecolor{ultramarine}{RGB}{0, 32, 96}
\definecolor{seablue}{RGB}{140, 212, 199}
\definecolor{lightyellow}{RGB}{255, 255, 178}
\definecolor{lightpurple}{RGB}{189, 186, 216}
\definecolor{lightred}{RGB}{250, 127, 115}
\definecolor{lightblue}{RGB}{127, 178, 212}
\definecolor{lightorange}{RGB}{255, 181, 97}
\definecolor{grassgreen}{RGB}{178, 222, 105}
\definecolor{lightpink}{RGB}{252, 204, 230}
\definecolor{lightgray}{RGB}{217, 217, 217}
\definecolor{darkpurple}{RGB}{188, 127, 188}
\definecolor{lightgreen}{RGB}{204, 235, 196}
\definecolor{honeyyellow}{RGB}{255, 237, 112}

\renewcommand\footnotetextcopyrightpermission[1]{} 
\setcopyright{none}

\acmDOI{}

\acmISBN{}

\acmPrice{}

\begin{document}
\title{Rediscovering Recurring Routing Results}
\ifisanon
\subtitle{IMC'25 submission \#84,  \pageref{page:last_body} pages body, \pageref{page:last_page} pages total}
\fi

\ifisanon
  \newcommand{\OurUniversity}{our university}
\else
  \newcommand{\OurUniversity}{USC}
\fi

\ifisanon

\author{Anonymous}

\else

\author{Xiao Song}
\affiliation{
	\institution{University of Southern California}
	\country{United States}
}

\email{songxiao@usc.edu}

\author{John Heidemann}
\affiliation{
	\institution{University of Southern California}
	\country{United States}
}
\email{johnh@isi.edu}

\fi

\begin{abstract}
Routing is central to networking performance,
 including: (1) latency in anycast services
 and websites served from multiple locations,
 (2) networking expenses and throughput in
 multi-homed enterprises,
 (3) the ability to keep traffic domestic
 when considering data sovereignty.
However, understanding and managing
\emph{how} routing affects these services is challenging.
Operators use Traffic Engineering (TE) with BGP
 to optimize network performance,
 but what they get is the result of \emph{all BGP policies}
 throughout the Internet,
 not just their local choices.
Our paper proposes \emph{Fenrir, a new system to rediscover recurring routing results}.
Fenrir can \emph{discover} changes in network routing,
 even when it happens multiple hops away from the observer.
Fenrir also provides new methods to \emph{quantify} the degree of routing change,
 and to \emph{identify} routing ``modes'' that may reappear.
Second, we show that Fenrir can be applied to many different problems:
 we use five instances of three different types of systems to illustrate the generalization:
 anycast catchments showing in a root DNS service,
 route optimization for two multi-homed enterprises,
 and website selection for two of the top-10 web services.
Each type requires different types of active measurements,
  data cleaning and weighting.
We demonstrate Fenrir's methods of detecting and quantifying change
 are helpful because they all face similar operational questions:
How much effect did traffic engineering have?
Did a third-party change alter my routing?
In either case, is the current routing new, or is it like a routing mode I saw before?
\end{abstract}

\maketitle

\section{Introduction}

Routing has been a core component of the Internet since its inception~\cite{Clark88a}.
Since the early 1990s,
  routing meant finding a good available path
 and quickly selecting an alternate after a routing failure.
As the Internet commercialized in the mid-1990s,
 route \emph{optimization} grew in importance~\cite{Rexford06a},
 often to minimize cost while considering performance.

\textbf{The Need to Understand Routing and Optimization:}
The growth of hypergiants (such as Google and Facebook~\cite{Labovitz10c})
 has made route optimization mandatory,
 since egress traffic can overwhelm individual links~\cite{Jain13a,Schlinker17a,Yap17a},
 and minimizing network latency is often a competitive advantage.
With their global backbones (for example,~\cite{Jain13a}),
 and extensive global peering mixing transit providers with public IXPs~\cite{Ager12a}
 and private interconnects~\cite{Oliveira08a}
 at hundreds of points-of-presence (PoPs)~\cite{Calder13a},
 route optimization has huge effects on hypergiant operational expenses
 and performance for their customers.
The effects of traffic shifts due to the
 growth and broad-peering hypergiants
 has been described as ``the flattening of the Internet'',
 a trend in Internet routing~\cite{Labovitz10c,Chiu15a}.

Routing also drives IP Anycast operations,
 determining \emph{Anycast catchments},
 which traffic goes to which physically distributed anycast sites.
Today, the broad use of IP Anycast by 
 DNS~\cite{Partridge93a,Hardie02a,Colitti06a} and CDNs~\cite{Schomp20a,Koch21a}
 to minimize latency~\cite{Schmidt17a} and manage DDoS~\cite{Moura16b,Rizvi22a}.

Routing's role in business competition prompts
 network operators to understand how routing affects their customers,
 and to 
 quickly detect routing changes.
Route optimization tools are commercially available,
 and hypergiants often deploy custom tools (for example,~\cite{Schlinker17a,Yap17a}).
Users of IP Anycast have developed special 
 tools to evaluate destinations~\cite{Woolf07a,Vries17b,Sommese20a}.
Companies (ThousandEyes, Cisco, Dyn) and and non-profits (RIPE)
 monitor routing as a service,
 often with thousands of distributed observers~\cite{Ripe15c}.
Finally,
 public routing data has been available 
 for 20 years~\cite{RouteViews00a,Ripe11a}.
 Interpreting this data and improving routing remain important research topics.

Despite its importance,
\emph{today, there is no good way to summarize routing for a service,
nor to differentiate between tiny changes and huge shifts}.
It is challenging to understand routing, even for a single destination,
	because no one organization has a global view of routing---routing
	emerges from combining the Internet's diverse routing policies 
	as determined by business and policy constraints.

Of course, service operators know about routing changes they initiate,
and their transit providers may notify them of changes in their immediate upstream.
But routing is a function of \emph{everyone in the Internet},
policy or operational changes multiple hops away can affect service performance.
One specific recent example is unexpected cable cuts in the Baltic Sea~\cite{Pancevski24a},
resulting in latency seen in some European networks~\cite{Aben24a}.
These changes were explained with a one-off research analysis,
but in general operators have \emph{no direct way} of detecting
how much routing changes beyond their immediate neighbors affect their reachability. 

\textbf{Contributions:}

The first contribution of our work is to present \emph{Fenrir,
	a new approach to rediscover recurring routing results}.
Fenrir takes inputs from measurement systems
	of distributed systems running across the Internet
	that are affected by routing,
	cleans and distills information into \emph{routing results} that summarize
	state of the service in terms of \emph{catchments}
	and relate to how networks access the service.
We quantify differences between vector pairs
  over the entire time series,
  and automatically cluster vectors to find
  mostly stable ``modes'' in routing results.
We visualize those differences
  to help operators reach actionable decisions.

Our second contribution is to
  show that \emph{Fenrir can apply to various network
  services} where routing plays a critical role.
We evaluate customer assignment to sites in IP anycast systems
  with the \broot DNS service.
We examine routing out a multi-homed enterprise, \EnterpriseUniversity,
	examining both immediate upstreams and our upstream's transit providers.
Finally, we examine Google's website,
	a top-10 website that is served by thousands of front-ends.
We show that Fenrir can also help make sense of
Google's DNS-based redirection, in addition to routing.
Fenrir applies to these different services,
  each poses unique 
  data collection and cleaning needs.

Although these services differ in detail,
  they are common in each,
  \emph{Fenrir helps service operators answer important operational questions}.
How quickly do catchments change?
Do routing results re-occur at later times?
Are two 80\% similar, or 40\%?
How often do third-party routing changes (perhaps multiple hops upstream)
  change catchments?
While we directly report about routing,
  network operators can combine routing with service-specific
  information such as traffic load or customer latency
  to quickly understand the implications of intentional and third-party
  routing changes.

The third contribution of this paper
  is to \emph{study these real-world systems
  to validate Fenrir accuracy,
  and to see what Fenrir shows about each}.
We validate Fenrir against ground truth from \broot operators (\autoref{sec:validation}),
	proving that Fenrir sees nearly all
	known routing changes, with recall of 1 and accuracy of 0.84.
We also show data that suggests that third-party routing changes
	have visible effects on service-specific catchments.

We examine what Fenrir finds in our three services of interest (\autoref{sec:results}).
Based on eight-month data of the multi-homed enterprise \OurUniversity,
	we show a significant routing change indicating that at most 90\% of
	catchments have changed.
By evaluating the five-year data of \broot,
	the routing is relatively stable;
	adding new anycast sites and removing old sites result in new modes.
Also, by comparing the routing results at the end of 2019 and the end of 2024,
	we observe that around 30\% of networks fall back to previous routing mode.
From two months of data for Google, 
	we confirm their aggressive policy for service evolution.
From one and half months of data for Wikipedia,
	we observe that one site was drained and later up again,
	however, the new routing result is only 80\%
	similar to the previous one.

\textbf{Data availability:}
Our paper uses a combination of publicly available data
  and new datasets we collect, as shown in \autoref{tab:datasets}.
In addition to existing public RIPE Atlas data,
  we will release 
  our enterprise and top-website datasets
  to researchers at no cost,
  with the publication of our paper.
  
\textbf{Ethics:}
Our paper studies network routing and its impact on
  public-facing services,
  so our data poses no privacy concerns to individuals.
We do collect data to and from millions of networks,
  but we have no knowledge of which networks are used by specific
  individuals,
  and we will not join our data with user-to-location mapping
  data, nor allow others to do so,
	as required by a usage agreement.

\section{Methodology}
	\label{sec:methodology}

\begin{table*}
	\resizebox{\textwidth}{!}{%
	\begin{tabular}{lll}
        \textbf{Step} & \textbf{Goal} & \textbf{How} \\
        Identifying subjects (\autoref{sec:data_collection}) &  Define target networks and destinations   &  Public datasets\\
        \hline
        Data collection (\autoref{sec:data_collection}) & Observe associations between networks and service cachements & Active probing \\
        \hline
        Data cleaning (\autoref{sec:data_cleaning}) & Remove bogus data; fill missing holes & Service-specific  \\
        \hline
        Comparison function (\autoref{sec:data_comparision})&  Direct pair-wise comparison among pairs and timeseries & Gower's distance \\
        Clustering (\autoref{sec:data_clustering}) & Find routing ``modes''  & Hierarchical Clustering\\
        \hline
        Quantification (\autoref{sec:data_visualization}) & Show changes and similarities &  Heatmap and transition matrix  \\
        \hline
        Performance (\autoref{sec:data_evaluation_performance}) & Evaluate performance of routing modes  &  Latency  \\
      \end{tabular}
      }
    \caption{Our steps to rediscover recurring routing results.}
    \label{tab:short_steps}
  \end{table*}

Fenir's process is summarized in \autoref{tab:short_steps}.
We actively collect data that shows how routing affects the observer.
After cleaning, it generates vectors that represent routing results.
We show how to compare vectors to quantify
	how similar routing is on different days,
	and to discover and visualize significant changes.

\subsection{Problem Statement}
\label{sec:problem_statement}

Routing establishes paths from \emph{global networks} to \emph{services}.

To capture how routing affects a service,
	we define \emph{catchments} as cut (in the sense of graph theory)
	between those networks and that service.
Our use is a generalization of how catchment is used in IP Anycast,
	where catchments are the list of server sites
	and reflect which client networks visit each site
	(and, in turn, from hydrography, where catchments are the drainage basins of rivers).

Our generalization considers
	not only catchments at the service,
	but also in the \emph{middle} of the network.
For example, for a multi-homed enterprise,
	catchments can represent the path upstream provider
	used to reach each destination network at each hop.
Understanding catchment evolution
	is important and challenging because
	they depend not only on routing
	policies at the source and destination,
but also by the \emph{routing policies of parties in the network}.

\begin{table*}
  \resizebox{\textwidth}{!}{%
  \begin{tabular}{lllllll}
  \textbf{case study} & \textbf{service} & \textbf{catchment} &  \textbf{network} & \textbf{dataset:} name & start & duration \\
    \hline
  anycast & DNS or anycasted services  & anycast sites & 5M IPv4 /24 blocks~\cite{Vries17b} & \broot/Verfploeter &  2019 & 5 years \\
     & DNS or anycasted services & anycast sites & 13k RIPE Atlas VP~\cite{Ripe15c}& \broot/Atlas & 2019 & 5 years \\
    \hline
  multi-homed enterprise & an enterprise  & upstream providers & 1.6 M IPv4 /24 blocks & \EnterpriseU/traceroute & 2024-08 & 8 months \\
    \hline
  top websites & a hypergiant website  & website instances & global networks & Google/EDNS-CS & 2024-02 & 2 months \\
   & a non-profit website  & website instances & global networks & Wiki/EDNS-CS & 2025-03 & 1.5 months \\
  \end{tabular}
  }
  \caption{Paper terms (network, catchment, service) and datasets
    used for
    three different systems.}
    \label{tab:terms}
    \label{tab:datasets}
  \end{table*}

We aim to summarize the catchment state in \emph{routing results}.
Mathematically, a vector contains all information about a service's catchments at a specific time.
It defines a particular routing result,
	and allow operators to determine
	what changes and if a routing result re-occurs.
We \emph{compare} vectors
	to quantify degrees of similarity or difference;
	operators would like to know if routing is ``80\% like last month''.
Operators often track user-relevant performance metrics,
	so a 20\% similar vector may require re-investigation
	of performance gains and losses,
	but if a route change shows a vector that is 95\% like last quarter,
	the operator will have confidence in what performance to expect
	even before a new investigation.

Routing affects Internet services and traffic in general,
  and we expect Fenrir can be applied to
 several problem domains.
\autoref{tab:terms} lists three classes of systems
  and six combinations of measurement methods.
We study 1) traffic routing from a multi-homed enterprise such as a university;
	2) anycast catchments, as are widely used by large DNS providers~\cite{Abley06a,Bellis10b} 
	and Content Delivery Networks~\cite{Alzoubi08a,Calder13a};
  3) websites for major Internet properties, like Google 
  or Facebook~\cite{Calder13a,He13a,Fan15a,Schlinker17a} and for non-profit organization, like Wikimedia that hosts Wikipedia.
All of these services share a common need to understand
  and influence routing.
Multi-homed enterprises seek to maximize performance for users
  while minimize traffic across peers which may be no-cost, 
  fixed cost, or metered.
DNS providers and CDNs strive to 
  lower latency~\cite{Liu07a,Bellis15a,Calder15a,Flavel15a,Schmidt17a,Li18a,Wei20a,Koch21a,Zhang21a,Rizvi24a}
  and manage traffic during large DDoS 
  events~\cite{Moura16b,Moura18b,Rizvi22a}.
Major Internet services seek to balance ingress and egress traffic,
  not overloading links and maximizing user 
  QoE (for example,~\cite{Schlinker17a}).
Organizations such as RIPE
  evaluate country-level Internet access~\cite{Ripe23a}
  with metrics such as AS-hegemony~\cite{Fontugne18a}

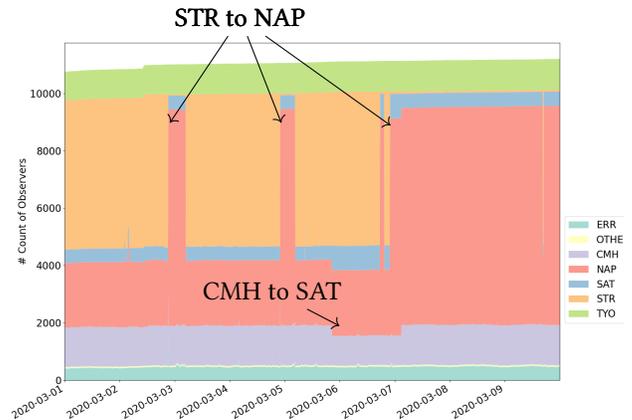
\begin{figure}
  \mbox{\begin{annotationimage}{width=1\linewidth,clip}{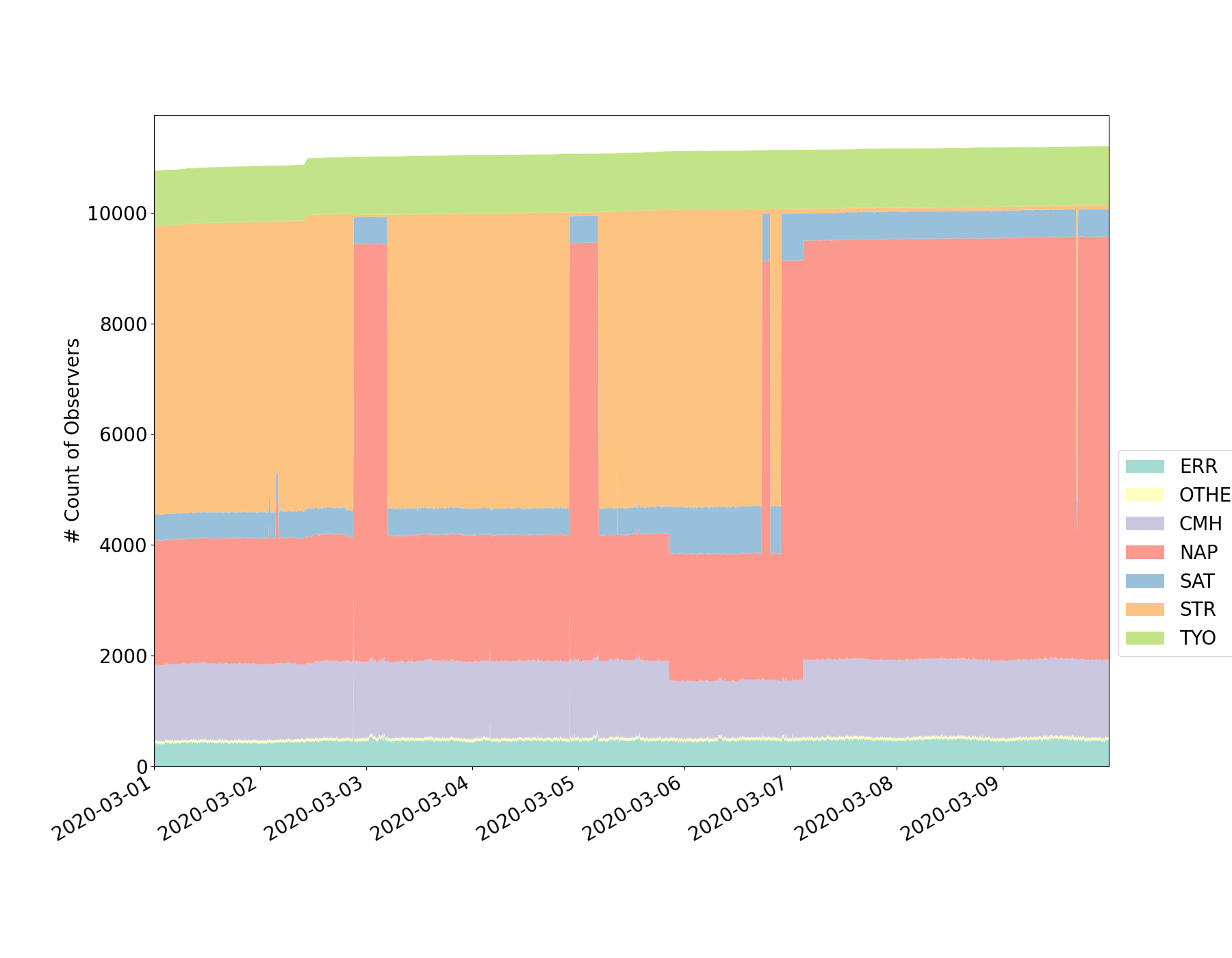}
    \path (0.4,0.93)node(x){STR to NAP}(0.28,0.7)node(y){};
    \draw[->,black,font=1pt] (x) -- (y);
    \path (0.4,0.93)node(x){STR to NAP}(0.47,0.7)node(y){};
    \draw[->,black,font=1pt] (x) -- (y);
    \path (0.4,0.93)node(x){STR to NAP}(0.65,0.7)node(y){};
    \draw[->,black,font=1pt] (x) -- (y);
    \path (0.45,0.38)node(x){CMH to SAT}(0.57,0.3)node(y){};
    \draw[->,black,font=1pt] (x) -- (y);
  \end{annotationimage}}
  \caption{Catchment sizes in \groot, as measured by counts of RIPE Atlas VPs (measurement id 10314).
    2020-03-01 to 2020-03-09.}
  \label{fig:g.root.2020030x.count}
%
%
\end{figure}

\textbf{A Case Study:}
As an example illustrating our goals,
  \autoref{fig:g.root.2020030x.count} shows
   ten days of anycast catchments for \groot. 
Each of the five different colors represents
  and how many observers are sent to each anycast site
  (the anycast catchment)
  for five sites (one is not observed) and error and other responses.

This visualization helps us detect operationally important changes.
Around midnight on 2020-03-03,
STR (in Stuttgart, Germany) almost completely drains,
  with its users shifting to NAP (in Naples, Italy).
This situation reverts 4.5\,h later.
This same mode happens again on 2020-03-05,
  and a third time, on 2020-03-07, it shifts again and remains through
  the end of observation.
With one site completely draining, we suspect these changes reflect
  maintenance activity by the service operator.

We also see a second, smaller shift of users from CMH to SAT
  for two days, starting on 2020-03-06.
This event could represent intentional route changes by the operator,
  or it could be changes resulting
  from some third-party network's routing policy.

Our goal is to assist operators
  in discovering these changes
  and understanding how
  affect service:
  how many users shifted after maintenance?
What was the performance of the different maps?
Were the shifts of users the same of different on different days?
If the secondary event was caused by a third party,
  were operators aware of it?
Did it affect performance for clients that kept the same catchment?

\subsection{Defining the Vector}
\label{sec:define_the_vector}

In \autoref{tab:terms}, our goal is to relate how networks
  use services, with cachements representing a cut through the network graph.
We formalize a vector at time $t$ as 
  $D(t)$, with one element for each of the $N$ networks.
Each element takes one of $|S|$ values,
  one for each possible service site in $S$,
  as $S$ is the set of service sites.
As an example, part of the vector for \autoref{fig:g.root.2020030x.count}
  on 2020-03-01 might be:
  \[
    D=[ \Vr{CMH}, \Vr{NAP}, \Vr{STR}, \Vr{STR}, \Vr{OTHER}, \Vr{SAT}, \Vr{ERR}, \dots ]
  \]
It means that the first network goes to CMH, the second network goes to NAP, the third and forth network go to STR, etc.

A one-hot representation is helpful mathematically,
  where we define a normalized vector,
  as the matrix $D_*(t)$, size $N \times |S|$,
  where each cell $D_*(t,n,s)$ is 1 if $D(t,n) = s$, otherwise 0.

We can summarize how many networks go to each site based on aggregating by site
  into an $|S|$-long vector $A(t)$.
  \[
    A(t,s) = \sum_{\forall n \in N} D_*(t,n,s)
  \]
Continuing our example from 2020-03-01 in  \autoref{fig:g.root.2020030x.count},
we get an aggregated vector $A$ for sites CMH, NAP, STR, NRT, SAT, HNL, err, and other.
\[
  A(\V{2020-03-01}) = [ 1350, 2200, 5200, 1000, 480, 10, 560, 50 ]
\]

\noindent Two days later on 2020-03-05, when STR mostly drains:
\[
  A(\V{2020-03-05})  = [ 1350, 7300, 1, 1000, 480, 10, 440, 50 ]
\]

\subsection{Data Collection}
	\label{sec:data_collection}

We use active measurements to determine the current routing result.
The exact protocol used varies by the service being studied,
  with the requirement that it be able to indicate
  the catchment each network is in.
Next, we describe measurement protocols for each service next,
  and then how we scale raw observations to normalize coverage.
\autoref{tab:datasets} lists our datasets and measurements.

\subsubsection{Anycast}
\label{sec:data_collection_anycast}

Anycast catchments are defined by the anycast site for each client to reach,
	as determined by BGP routing.
We use two methods to map anycast catchments:
  RIPE Atlas observes which DNS root server they reach
  from each of their about 10k vantage points (VP)~\cite{Ripe15a}.
RIPE Atlas has archived such measurements
  for more than a decade.
Atlas uses standard DNS mechanisms
  (CHAOS queries to hostname.bind or NSID~\cite{Woolf07a})
  to identify a per-server identifier.
We then map these organization-specific identifiers
  to determine specific anycast sites, following prior work~\cite{Fan13a}.
\autoref{fig:g.root.2020030x.count} shows an example of catchments at G-Root
  using this method.
  
Alternatively,
  we measure anycast catchments from an anycast system using Verfploeter~\cite{Vries17b}.
Verfploeter requires the involvement of the anycast system,
  and the \broot operators have shared about three years of data with us.
Verfploeter provides broader coverage than Atlas,
  associating 5M networks with their \broot destinations.
Verfploeter gets this coverage, pinging targets in millions of networks
  and watching which catchment the reply goes to
to determine the target's current catchment.
\autoref{fig:broot_2019_2025} shows Verfploeter data for \broot.

\subsubsection{Network connectivity}
\label{sec:data_collection_network_connectivity}

Cachements in network connectivity are defined by what upstreams
or transit providers one uses to reach destination networks.
We map enterprise upstream catchments
	by taking traceroutes out of the enterprise
	to all routable network prefixes.
Then, we look at a reasonable distance outside campus, but not infinitely far away.
RIPE maps country-level transits with control-plane observations:
	they observe routing from RIPE RIS route collectors,
	and look through the AS paths in the BGP routing table
	to all IP blocks in the country under study.

Unlike anycast catchments, which are defined by the anycast site
	each network is assigned to,
	network connectivity can consider entire paths.
While enterprises are usually concerned about their immediate upstreams,
	and RIPE country-level analysis studies Tier-1 transit providers
	out of the country,
	in both cases, one can adjust the ``focus'' of the study to consider
	more or fewer hops.

We use scamper~\cite{Luckie10a} to take traceroute from one server at \EnterpriseU,
	measuring the first 10 hops
	to selected routable addresses based on a hitlist~\cite{Fan13a} of every /24 blocks.
We cover 1.6M /24 networks by comparing hitlist and routable blocks
obtained from the BGP table of RouteViews.
We terminate Traceroute when it exceeds the 10th hop.
It takes around 8 hours to complete a full list scan.
The probing rate is 550 packets per second.
We assume the underlying routing does not change rapidly,
	keeping the probing rate low to reduce the stress
	on the first hop while being able to capture the changes.

\autoref{fig:heatmap_enterprise_hop_3} 
  is an example of routing out of one multi-homed enterprise,
  \OurUniversity.

\subsubsection{Top Websites}
	\label{sec:data_collection_top_website}

Like anycast, catchments in top websites are defined
	as the site that serves the request for each network that originates a query.
We measure top websites by making queries to URLs for the front page of a website.
Most websites use DNS-based load-balancers that select
	specific servers near the target,
	so website evaluation involves both DNS and HTTP queries from different locations.

We want to understand why website front-ends are chosen
	for all potential global observers.
Since no one has observers in all networks,
	we use the EDNS/Client-Subnet approach from Calder et al~\cite{Calder13a}.
To evaluate the catchment from a given observer $N$,
	we make a DNS lookup for the hostname in the website URL
	given the prefix $N$ as the client-subnet option.
For websites that use DNS-based server selection,
	if the DNS recursive resolver passes Client-Subnet through,
	we can map global catchments using a single physical observer.

We studied two targets using this method:
  Google, with thousands of front-ends aggressive churn,
  and Wikipedia, with seven global sites and more modest changes.
\autoref{fig:EDNS.20240217.20240421.heatmap} shows Google rapidly changing routing, and 
	\autoref{fig:edns_wiki_heatmap} shows a significant routing change at Wikipedia on 2025-03-19.

\subsection{Data Cleaning}
	\label{sec:data_cleaning}

Although active measurements provide input to our algorithms,
such raw observations often have errors or gaps,
	therefore we clean raw observations in three ways.

\textbf{Remove Incorrect Data:}
Some measurements contain incorrect data.
The details vary by service,
but when we identify clearly incorrect data, we discard it.

\textbf{Remove Micro-catchments:}
Some services
  have many sites,
  but not all are equally important.
We therefore identify \emph{micro-catchments}, 
	which are defined to be sites that are responsible for few newtorks, 
	as determined by what we observe in our observation system.
Removing micro-catchments reduce the complexity of identifying routing results.
It helps us focus on large catchments that would significantly affect routing.

Examples of micro-catchments occur in
	anycast and enterprise routing.
In anycast, local-only sites serve only a single AS and its customers,
	not the general Internet.
If no VPs are observed from within that AS,
	catchments for local sites will be of zero size.
Second, in enterprise routing,
	we are most concerned about how the enterprise reaches the world.
However, the enterprise may have routes to a few local network prefixes,
	such as networks within it.
We can filter out such networks if desired.

\textbf{Interpolate Missing Data:}
One-shot active measurements often miss data
due to random packet loss.
Even with retries, temporary network anomalies can result in
data gaps.
Missing values are problematic as they may
skew the data towards a lower availability~\cite{Heidemann08c}.
We fill gaps by interpolating data from
recent successful observations.
The details of what kind of gaps occur and how we fill them vary
by service.

Verfploeter sends queries from a central site; if the query target is temporarily unavailable,
that block will have no known catchment.
We fill missing anycast observations by replicating the most recent
successful observation.
The same cleaning strategy applies to top website measurements
with EDNS/CS as well.

Measurements of enterprise routing with traceroute
provide redundancy as each hop is queried multiple times,
but intermediate traceroute hops with private addresses or that filter pings
will not provide results.
Since traceroutes are often successful at some hops even if unsuccessful at others,
we use this spatial redundancy and propagate the nearest viable hop
to fill a traceroute gap.

We adopt nearest neighbor imputation to fill up the hole of missing values.
Specifically,
we address the losses of consecutive probes
by identifying two positive responses separated by a series of non-responses.
We then fill this gap with positive responses from our nearest neighbors.
For example, a list of consecutive probes is denoted by $ [1,\dots, n]$.
Assume a series of consecutive probes $[k,\dots, k+i]$ is missing.
To fill up missing holes,
all probes in $[k, \dots, k+i/2]$ are replaced by $[k-1]$ and
all probes in $[k+i/2, \dots, k+i]$ are replaced by $[k+i+1]$.
We put a limit (up to 3 observations away) on $k$ to how far
we can interpolate the missing value to avoid changing vectors.

\subsection{Weighting}
	\label{sec:weighting}

Once cleaned, our measurements provide a preliminary list of
of catchments that associate networks with services.
While true, this list may not represent metrics that are most operationally
important---each
observation indicates what that VP sees.
What operators care about is what that VP \emph{represents}---how many users are there,
or how much traffic is sent from that location.
We therefore adjust raw observations with a service-specific
\emph{weighting factor}
to represent metrics of operational interest, like IP addresses, traffic, or users.
We therefore assign a weight vector $D_w(t)$
that parallels the vector, showing how ``important'' each component is.
A default weight vector might be all 1, showing each observation
is equivalent,
but alternatives might weigh observers differently,
considering the number of addresses, users, or traffic.

For anycast, we can weight results using networks and traffic estimates.
Observers with RIPE Atlas and Verfploeter may not be uniformly distributed
across the IPv4 address space,
so, as our first step, we can weight each observation by how many
networks it represents.
Thus, if we have only one Atlas VP or Verfploeter response from a /16 prefix,
we can count that as 256 /24 blocks rather than just one.
We make the assumption that addresses in the same /24 blocks
are located in the same or nearby physical location.

For services that collect and summarize historical traffic or users
we can go one step further and weight blocks by the amount
of historical traffic they have sent,
or users they have.
Verfploeter compared addresses and traffic for DNS~\cite{Vries17b}.
Since network occupancy and user traffic demands vary considerably,
so can these estimates.

Enterprise and country-level routing also benefits from weighting
by IP addresses and traffic estimates.

Top websites should be weighted by the number of users in each
network.
Although we do not have access to user distributes and their traffic
demands, certainly websites have this information.

\subsection{Analysis: Comparion and Clustering}

Given a vector $D$ composed of clean data and weights ($D_w$),
  we next describe how to compare and cluster vectors to discover trends.

\subsubsection{Pairwise Comparision}
        \label{sec:data_comparision}

We first define a pairwise comparision function to judge how similar
  any two vectors are.

For pairwise comparion, we adopt Gower's Distance~\cite{gower1985properties} between all $N$ elements.
When comparing elements at times $t$ and $t'$ match if they are part of the same catchment:

\[
  M(t,t',n) = \begin{cases*}
    1,              & if $D(t, n) = D(t', n) \wedge D(t, n) \neq \Vr{unknown}$,\\
    0,                    & otherwise.
  \end{cases*}
\]

And the normalized, weighted similarity $\Phi(t,t')$ of Gower's Distance is:

\[
  \Phi(t,t') = \frac{\sum_{n \in N} M(t, t', n)  D_w(n)}{\sum_{n \in N} D_w(n)}
\]

$\Phi$ has a physical meaning: it is the fraction (from 0 to 1.0)
of networks that are the
same between two vectors.
We plot distances between all-pairs combinations of vectors.
using a method we describe below \autoref{sec:data_visualization},
providing heatmaps such as
\autoref{fig:EDNS.20240217.20240421.heatmap}.

The test for ``unknown'' in $M(t, t', n)$
treats all cases where we do not know network $n$'s catchment
as changing.
This assumption is pessimistic and reduces $\Phi$ for services where
observations are imperfect.
Measurements are frequently imperfect for Verfploeter measurements
of catchments since
Verfploeter requires a ping response to confirm catchment,
but predicting a responsive IP address in a target network
employing dynamic address assignment is probabilistic.
Since Verfploeter reports unknown for about half of it is 5M target networks,
a stable catchments will show $\Phi$ values between 0.5 and 0.6,
rather than near 1.0.
(As ongoing work, we plan to remove unknown networks from consideration
when computing $\Phi$, so it will define similarity of \emph{known} networks.)

\subsubsection{Clustering}
	\label{sec:data_clustering}

The comparison allows us to visualize the similarity of different vectors,
but to automatically discover similar groups
we use Hierarchical Agglomerative Clustering (HAC),
a method of unsupervised machine learning~\cite{Sibson73a}.
HAC merges vectors into clusters $\mathbf{T}_c$, the set of times
where vectors $D(t) \forall t \in \mathbf{T}_c$ are similar,
using a bottom-up strategy.

HAC will continue clustering until all vectors are in one cluster,
so we define a distance threshold to indicate when clusters are different
enough to be considered separate.

We determine the Distance Threshold adaptively.
One challenge is that entities are operated individually
and have a different number of target networks and destinations.
To find the best distance threshold for different entities,
we loop over a range of distance threshold $[0,1]$ with step $0.01$ and
construct a new HAC model with the distance threshold.
We choose the first HAC model with less than $15$ clusters with at least $2$ valid observations.
In practice,
we note that the number of clusters converges quickly when the distance threshold increases.
One can tune the parameters to fit its needs better.

\begin{table*}
\subfloat[Large shift from STR to NAP, from 21:56 to 22:00.]{
  \label{tab:trans_mat_example}
  \resizebox{\columnwidth}{!}{%
  \begin{tabular}{r@{\hspace{0.5ex}}r|rrrrrrrr}
      & & \multicolumn{8}{c}{subsequent state} \\
  &      & CMH & NAP & STR & NRT & SAT & HNL & err  & oth\\ \hline
 \parbox[t]{2mm}{\multirow{8}{*}{\rotatebox[origin=c]{90}{initial state}}} 
  & CMH & 1352 & 0    & 0    & 1    & 0    & 0    & 16   & 0     \\ 
  & NAP  & 0    & 1939 & 0    & 1    & 0    & 0    & 272  & 0     \\
  & STR  & 0    & \LGreenCell 3097 & 625  & 4    & 0    & 0    & \LRedCell 1542 & 0     \\
  & NRT  & 1    & 2    & 0    & 985  & 0    & 0    & 30   & 0     \\
  & SAT  & 1    & 0    & 0    & 1    & 472  & 0    & 2    & 0     \\
  & HNL  & 0    & 0    & 0    & 0    & 0    & 12   & 1    & 0     \\
  & err   & 17   & 45   & 15   & 14   & 7    & 0    & 309  & 0     \\
  & oth & 0    & 0    & 0    & 0    & 0    & 0    & 1    & 46    \\ 
  \end{tabular}
  }
  } \quad
\subfloat[Drain of STR completes, from 22:00 to 22:04.]{
  \label{tab:trans_mat_example_sec}
  \resizebox{\columnwidth}{!}{%
  \begin{tabular}{r@{\hspace{0.5ex}}r|rrrrrrrr}
      & & \multicolumn{8}{c}{subsequent state} \\
  &      & CMH & NAP & STR & NRT & SAT & HNL & err  & oth\\ \hline
\parbox[t]{2mm}{\multirow{8}{*}{\rotatebox[origin=c]{90}{initial state}}}
  & CMH & 1344 & 0    & 0    & 1    & 1    & 0    & 20   & 0     \\ 
  & NAP  & 0    & 4987 & 0    & 0    & 0    & 0    & 87  & 0     \\
  & STR  & 0    & 636 & \YellowCell 2  & 0    & 0    & 0    & 11 & 0     \\
  & NRT  & 1    & 0    & 0    & 993 & 0    & 0    & 8   & 0     \\
  & SAT  & 0    & 0    & 0    & 0    & 473  & 0    & 7    & 0     \\
  & HNL  & 2    & 24    & 0    & 0    & 0    & 12   & 0    & 0     \\
  & err   & 17   & \LGreenCell 1801   & 0   & 35   & 2    & 1   & 317  & 1    \\
  & oth & 0    & 0    & 0    & 0    & 0    & 0    & 0    & 45    \\ 
  \end{tabular}
  }
  }
  \caption{Transition matrices for G-Root on 2024-03-04.  Sites are airports, plus error and other states.
   }
  \label{tab:trans_mat_both}
\end{table*}

%

\subsection{Quantifying Differences Between Vectors}
	\label{sec:data_visualization}

While a vector summarizes the current routing result,
  and our normalized Grower's distance compares any two vectors,
  operators need to compare \emph{many} vectors
  to highlight periods of routing
  stability and the degree of change.

We summarize routing over time by comparing all pairwise vectors
  as a gray-scale heatmap,
  such as \autoref{fig:EDNS.20240217.20240421.heatmap}.
Heatmaps make it easy to identify blocks of similar routing results
  as high-similarity (dark-shaded) triangles,
  and changes as discontinuities in shading.

We sometimes augment heatmaps with a 
  \emph{transition matrix} to compare two vectors
  and summarize how they are different.
A transition matrix $T$ is an $|S|\times|S|$ matrix,
  where each $T(t,t',s,s')$ element evaluates how many networks
  were in state $s$ in vector $D(t)$ and are in state $s'$
  in vector $D(t')$.

\[
T(t,t',s,s') = \sum_{n \in N} 1 \textrm{\ if\ } (D(t,n) = s \wedge D(t',n) = s')
\]

Let $\Psi_{t,t'}$ to be a transition matrix 
  of $N$ target networks and $V$ destinations between time $t$ and time $t'$.
$\Psi_{t,t'}[c_v, c_u]$ denotes the number of target networks leaving from destinations $c_v$ 
  to destination $c_u$ from time $t$ to time $t'$,
  $ t,t' \in T$, $c_v, c_u \in V$.

For a completely quiescent network, $T(t,t')$ will be a diagonal
matrix equal both to $A(t)$ and $A(t')$.
However, when networks change the catchment between the two vectors,
they contribute to cells off the diagonal.
In \autoref{tab:trans_mat_example},
we see that 3097 networks
move from STR to NAP (in yellow),
and 1542 move from STR to the error state (they get no replies from any site).
In the next four minutes,
\autoref{tab:trans_mat_example_sec},
we see the transition completed and STR nearly completely drains,
with most sites moving from error to NAP.

\subsection{From Similarity to Performance Differences of Vectors}
	\label{sec:data_evaluation_performance}

While heatmap identifies regions of similarity
and
transition matrix specifies how any two vectors are different,
operators care about \emph{user relevant metrics},
not just how many routes changed.
\emph{Latency} is an important metric,
so we would like to estimate \emph{latency differences between pairs of vectors}

We therefore factor latency into our vectors
to allow operators to estimate the effects a routing change has on latency
and if they should investigate further and take some response.
We compute mean overall latency
by measuring latency for each network and weighting that value by pre-defined matrix~\autoref{sec:weighting}.
As with data collection, we use different data sources to estimate latency for each service.

\subsubsection{Anycast.}
\label{sec:latency_anycast}
Many measurement systems do probing to collect RTT toward the anycast sites.
For instance,
RIPE Atlas takes data from its 13k VPs
to all Root DNS with built-in measurements (\autoref{sec:data_collection_anycast}).
Each response contains RTT, representing the latency from the end host to the DNS anycast site.

\subsubsection{Network connectivity.}
\label{sec:latency_network_connectivity}
Measuring latency experienced by multi-homed enterprises is tricky,
as collecting end-to-end RTT requires scanning the whole Internet from the enterprise network.
In this case, we use Trinocular~\cite{Quan13a}, an existing Internet outage detection system operating since 2013.
One of Trinocular sites sits inside \OurUniversity; it scans approximately 5.2 million /24 IPv4 address blocks using ICMP echo request messages.
Each site probes between 1 and 16 targets per block every 11 minutes,
selecting targets from a fixed pseudorandom list updated quarterly.

Both data sources provide frequently updated RTT from VPs to the measured network,
with at least one site controlled by us,
satisfying our needs and without running extra measurements.

\section{Validation}
  \label{sec:validation}

To validate our work, 
	we compare changes detected by Fenrir with the ground truth: 
	verified operational data provided by the service operator.

\textbf{Ground truth:}
We assess Fenrir’s ability to rediscover recurring routing results
	by comparing its detected changes in \broot catchments
	against operator-provided ground truth over a four-month period beginning on 2023-03-01. 
Fenrir identifies events in the \broot/Atlas dataset 
	by examining transitions in vector matrices every four minutes. 
Many operator-reported events are short-lived, often lasting only tens of minutes.
They would not appear in 
	datasets collected at lower temporal resolution, 
	such as the daily \broot/Verfploeter data.

The ground truth is derived from \broot operator maintenance logs. 
Across this period, we observe 98 recorded maintenance entries. 
Because some maintenance activities are the combination of several events, 
	where some are externally visible and others not,
	we group events occurring within ten minutes and performed by the same operator into 56 event groups. 
We classify these as either invisible (internal changes with no observable external effect) or external (changes such as site drains or traffic-engineering adjustments that affect catchments).
As an example, an invisible event would be taking one of several replicated internal servers offline temporarily,
	where service automatically transfers to other active servers. 
A site drain indicates an intentional, 
	temporary removal of the site from anycast so that the router or firewall can be changed without query loss.
raffic engineering denotes routing adjustments that preserve reachability but shift catchments.
This operator-derived ground truth therefore captures only expected \broot routing changes.

\begin{table}

\begin{tabular}{lr@{\hspace{1ex}}lr@{\hspace{1ex}}l}
		 & \multicolumn{4}{c}{\textbf{Fenrir observations}} \\
\textbf{ground truth}  & \multicolumn{2}{c}{\textbf{detected in F.}} & \multicolumn{2}{c}{\textbf{not detected}} \\
  \hline
all logged events & \multicolumn{4}{c}{56 (98 before grouping)} \\
\quad  external &           \DGreenCell  19 & \DGreenCell (TP)  &  \LRedCell 0 &\LRedCell (FN) \\
\qquad    site drain &  	\quad \DGreenCell	17     &   & \quad \LRedCell 0 & \\
\qquad    traffic engineering &  \quad \DGreenCell  	2      &  &  \quad \LRedCell 0  &\\
\quad  internal only &       \YellowCell 8 &  \YellowCell (FP?*)  &  \LGreenCell 29 & \LGreenCell (TN) \\

\hline
\textbf{external changes?} &                   10 &  (*) &    --- &  \\ 
\end{tabular}
%
%
\caption{Evaluation of ground truth changes against Fenrir-visible changes
  for \broot/Atlas.}
	\label{tab:broot_confusion_matrix}
\end{table}

\textbf{Comparison:}
\autoref{tab:broot_confusion_matrix}
  compares ground truth (rows) against Fenrir observations (columns).
Our first conclusion is that
\emph{Fenir has good accuracy 0.86} ($(\V{TP}+\V{TN})/\V{all})$)
detecting most events,
with 19 true positives and 29 true negatives.

\emph{Fenir has perfect recall of 1.0} ($\V{TP}/(\V{TP}+\V{FN})$),
missing no events.
Our first examination of the data showed one false negative,
but confirmation of the operators revealed that that event
did have maintenance, but without draining the site,
making it an event that should be invisible to external detection.

Consideration of false positive events requires more care
because \emph{not all actual routing changes events are in ground truth}.
As presented,
Fenir precision is moderate at 0.70 ($\V{TP}/(\V{TP}+\V{FP})$),
because a direct comparison of ground truth against Fenrir's predictions
shows eight false positives.
However, we need to consider that Fenrir detects not only
ground truth events, but \emph{also} events that caused by routing
changes by third parties, such as transit providers.
Such changes will not appear in ground truth data, which considers only
operator-initiated changes.

In fact, \emph{detection of third party changes is Fenrir's design goal}
so this seemingly imperfect precision relative to known maintenance events
likely instead represents
\emph{10 third-party routing changes unknown to the anycast operator
and therefore of their interest---new visibility provided by Fenrir}.

Some evidence to support this hypothesis is that we also see
10 events in Fenrir (marked with (*))
that have no reflection in operator logs.
As ongoing work, we are currently looking at historic routing data
to find explicit evidence for such third-party events.

\section{Results}
	\label{sec:results}

We next examine how Fenrir can discover events
in each of our target systems listed in \autoref{tab:terms}:
multi-homed enterprises, anycast, and top websites.

To measure multi-homed enterprises,
we study the routing behaviors of the upstream ISP of \EnterpriseUniversity (\EnterpriseU).
We describe our approach to collect data at \autoref{sec:data_collection_network_connectivity}.
We have also observed a second enterprise for 10 months,
but thus far, we have not seen significant routing changes.

As an example system using anycast,
we measure \broot,
since we can get ground truth about routing changes
from its operators.
We measure it with one data source,
Verfploeter ~\cite{Vries17b},
from the \broot operators.
Unlike data collected by RIPE Atlas as used in \autoref{sec:validation} Validation,
which is run independently,
Verfploeter must be run by the anycast service,
but it provides data for the catchments selected by about 6.1M global networks.
We explain the details at \autoref{sec:data_collection_anycast}.

We measure two top-10 websites:
First, Google home page, served from thousands of sites~\cite{Calder13a},
and also Wikipeida's home page, served from 7 locations~\cite{WikimediaGlobalTraffic}.
In each case we use
EDNS/client-subnet
to simulate requests originating from 5M routable prefixes,
as described in \autoref{sec:data_collection_top_website}.

\subsection{Changes Upstream from an Enterprise}
    \label{sec:results_enterprise}

\begin{figure}

  \subfloat[Fraction of each catchment. ARN A: Academic Regional Network A.
	ARN B: Academic Regional Network B.
	ANN: Academic National Network.
 	HE: Hurricane Electric. ]{
    \label{fig:stackplot_enterprise_hop_3}
    \begin{tabular}{l}
    \imagelabelset{coordinate label back = none, coordinate label text= black, coordinate label font = \small}
    \begin{annotationimage}{width=1\linewidth,clip}{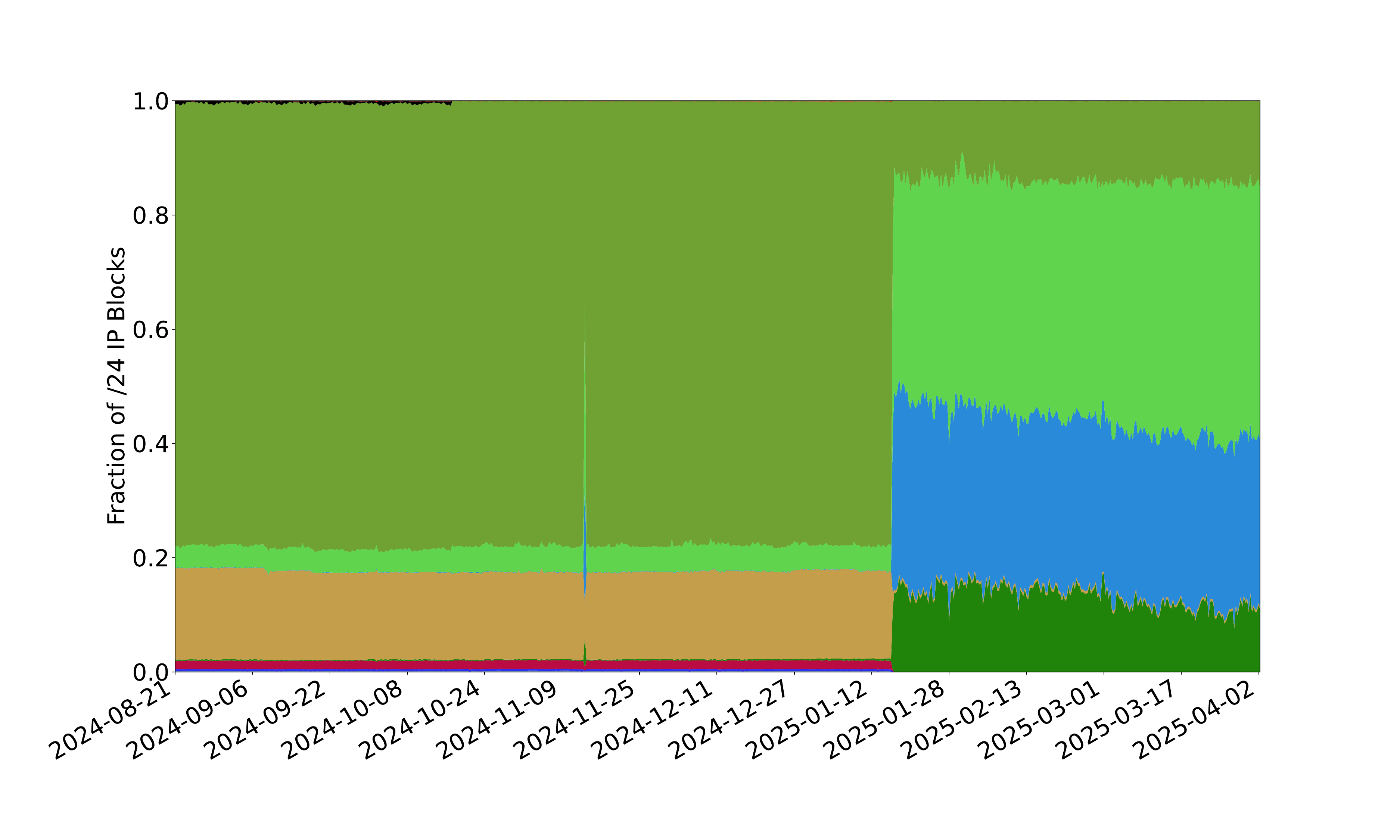}
      \draw[coordinate label = {ARN A at (0.3,0.7)}];
      \draw[coordinate label = {ANN at (0.3,0.25)}];
      \draw[coordinate label = {ARN B at (0.75,0.7)}];
	  \draw[coordinate label = {NTT at (0.75,0.4)}];
	  \draw[coordinate label = {HE at (0.75,0.25)}];
    \end{annotationimage}
  \end{tabular}
  }


  \subfloat[Heatmap of pairwise comparisions ($\Phi(t,t')$).]{
    \label{fig:heatmap_enterprise_hop_3}
    \begin{tabular}{l}
    \imagelabelset{coordinate label back = none, coordinate label text= black, coordinate label font = \small}
    \begin{annotationimage}{width=0.9\linewidth,clip}{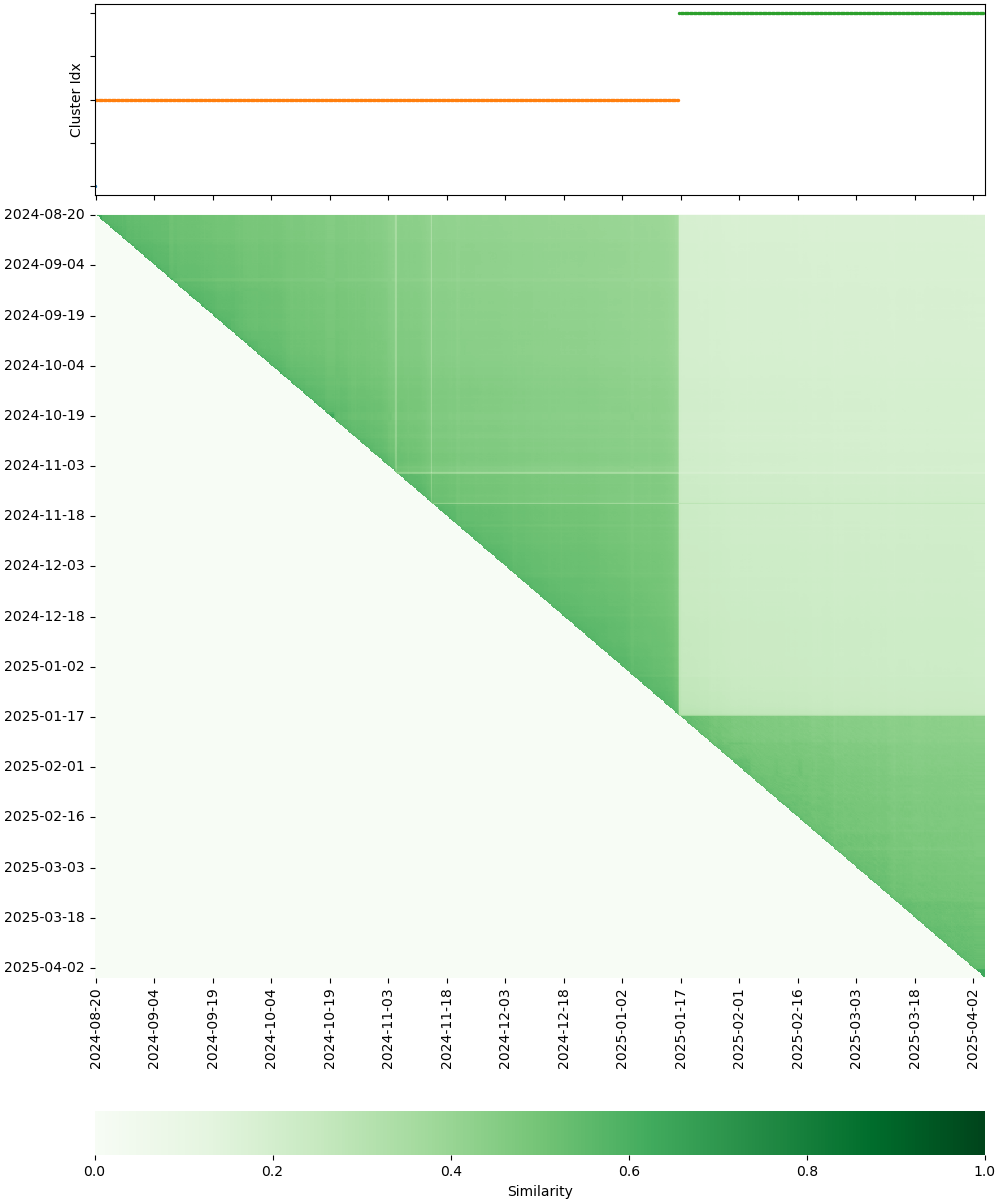}
      \draw[coordinate label = {(i) at (0.52,0.89)}];
      \draw[coordinate label = {(ii) at (0.9,0.95)}];
    \end{annotationimage}

  \end{tabular}
  }
  \caption{Enterprise catchments from 2024-08 to 2025-04 at hop 3. 
	The routing change happens on 2025-01-16. Dataset: \EnterpriseU/traceroute.} 
  \label{fig:enterprise_hop_3}
\end{figure}
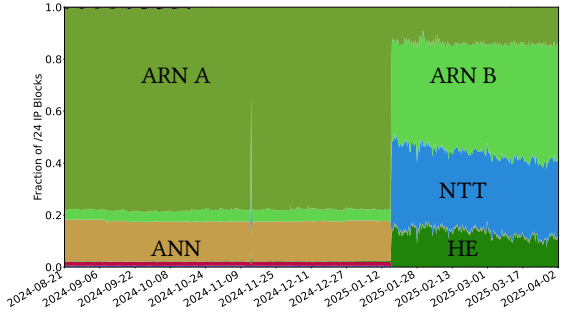
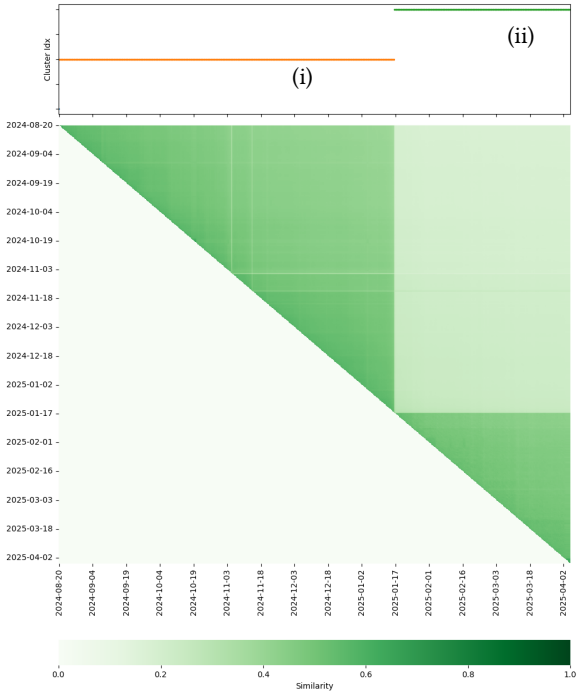

Many large enterprises are \emph{multi-homed},
connecting to multiple transit providers
and other peers
to send traffic to thousands of users
to the global Internet.
Large enterprises often
peer directly with important services, like cloud providers or CDNs,
or nearby peers and regional networks.
University, for example, benefit from
academic-networks (such as Internet2, GEANT, and ESNet)
that are often both high-speed and low-cost.
Traffic engineering and optimization can have a large impact
on cost and performance.
Such enterprises can optimize routing for cost and latency.

While an enterprise will directly monitor traffic at their borders,
  knows about their routing changes,
  and is likely informed about routing changes in their transit providers,
  \emph{they are often unaware of changes that happen elsewhere in the Internet}.
Yet such changes can be significant---unexpected cable cuts in the Baltic~\cite{Pancevski24a}
  resulted in latency changes in Europe~\cite{Aben24a},
  just one recent example of a change outside ones immediate peers
  with performance impact.

Fenrir has the potential to monitor and quantify routing changes
  that affect the path to any network.
We next show how Fenrir analysis
  can digest traceroutes to millions of networks (as described in \autoref{sec:data_collection_network_connectivity})
  to quantify changes to an enterprise's routing cone.

\subsubsection{Changes in Routing Modes:}
Here, we show how many modes are detected, how different they are, and why.

We first consider where traffic is, three hops away from the enterprise.
We use data covering eight months, starting 2024-08, for \EnterpriseUniversity.

\autoref{fig:heatmap_enterprise_hop_3}
  shows Fenir heatmap
  with two strong routing modes:
  mode (i) with $\Phi$ in [0.31, 0.65] and mode (ii) with $\Phi$ in [0.42, 0.57]).
	seperated by 2025-01-16.
Comparing $\Phi$ before and after the 2025-01-16 transition
  gives $\PhiModes{i}{ii} = [0.11, 0.48]$, a huge routing change.

While Fenrir's heatmap detects this change,
  we turn to the stack plot in
  \autoref{fig:stackplot_enterprise_hop_3},
Here
  with each colored region counting the number of destination networks
  ($A(t)$ from \autoref{sec:define_the_vector}, couting global /24 prefixes)
  that are handled by a given transit network.

We see that in 2024,
	almost all networks were served by \EnterpriseCenic and \EnterpriseInternetTwo.
There was a major routing change on 2025-01-06,
  80\% of networks are now served by
	\EnterpriseLN , 
	\EnterpriseNTT, 
	and \EnterpriseHE, 
  while \EnterpriseInternetTwo vanishes in hop 3.

We can also visualize the enterprise routing cone as a Sankey diagram,
  to see what networks traffic uses 2, 3, and 4 hops away.
Due to page constraints,
  we show before- and after-change Sankey diagrams
 in an appendix as
	\autoref{fig:sankey_enterprise_before} and \autoref{fig:sankey_enterprise_after}.

We believe this dramatic change was a network reconfiguration
  done by \EnterpriseU operators.
However, we were able to discover it via Fenrir heatmaps,
  and use Fenrir's data to visualize it with stack plots and Sankey diagrams.
For routing changes that occur multiple hops away, like the Baltic cable cuts,
  Fenrir's heatmap would be the only way an enterprise's network operators
  would be aware of the size of a change.

\subsection{Anycast Changes at \broot}
	\label{sec:result_b_root}

\begin{figure}

\subfloat[Fraction of each catchment.]{ 
	\label{fig:verfploeter_broot_stackplot_20190901.20250501}
  \begin{tabular}{l}
    \includegraphics[width=1\linewidth,clip]{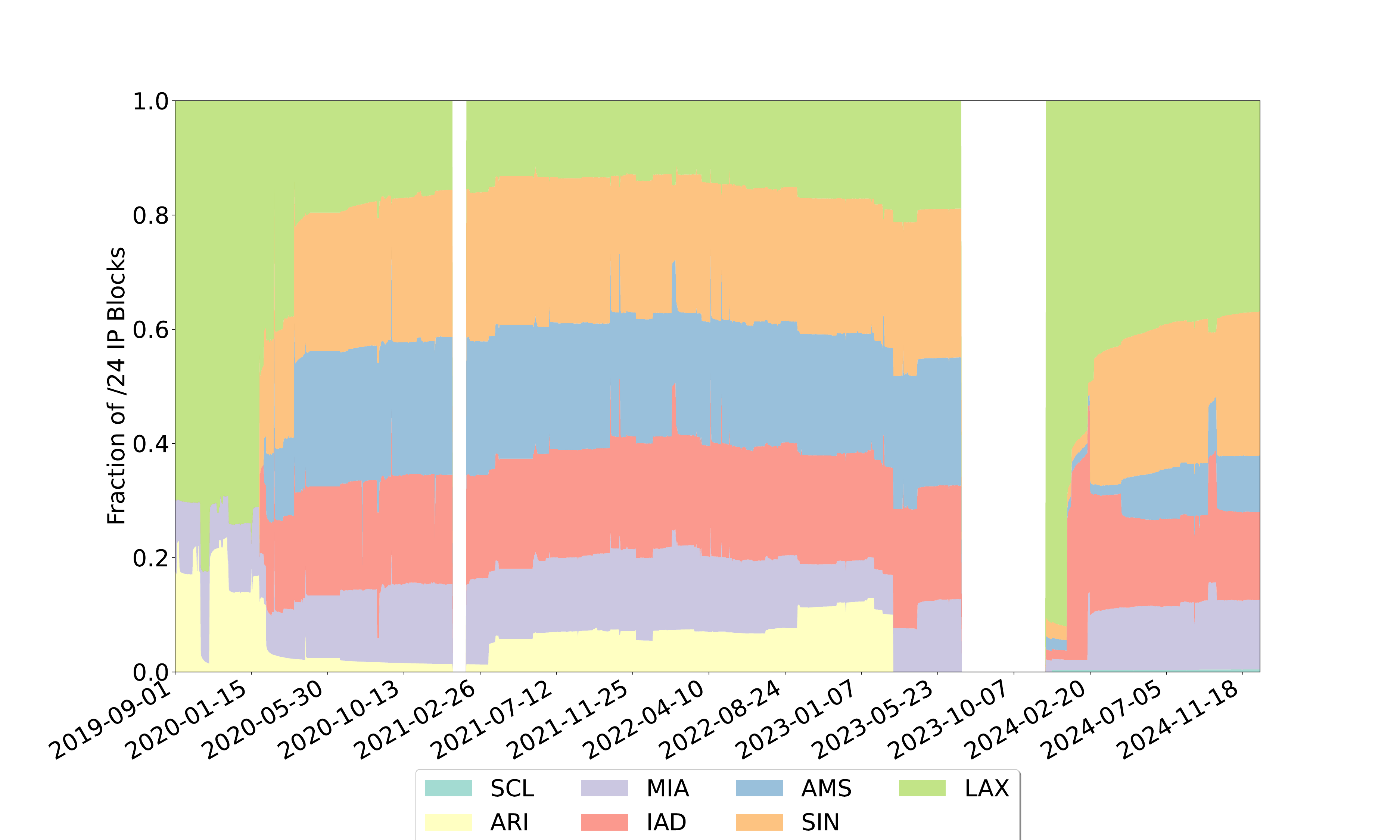}
\end{tabular}
}

\subfloat[Heatmap of pairwise comparisions ($\Phi(t,t')$).]{
 \label{fig:verfploeter_broot_heatmap_20190901.20250501}
  \begin{tabular}{l}
    \imagelabelset{coordinate label back = none, coordinate label text= black, coordinate label font = \small}
    \begin{annotationimage}{width=0.87\linewidth,clip}{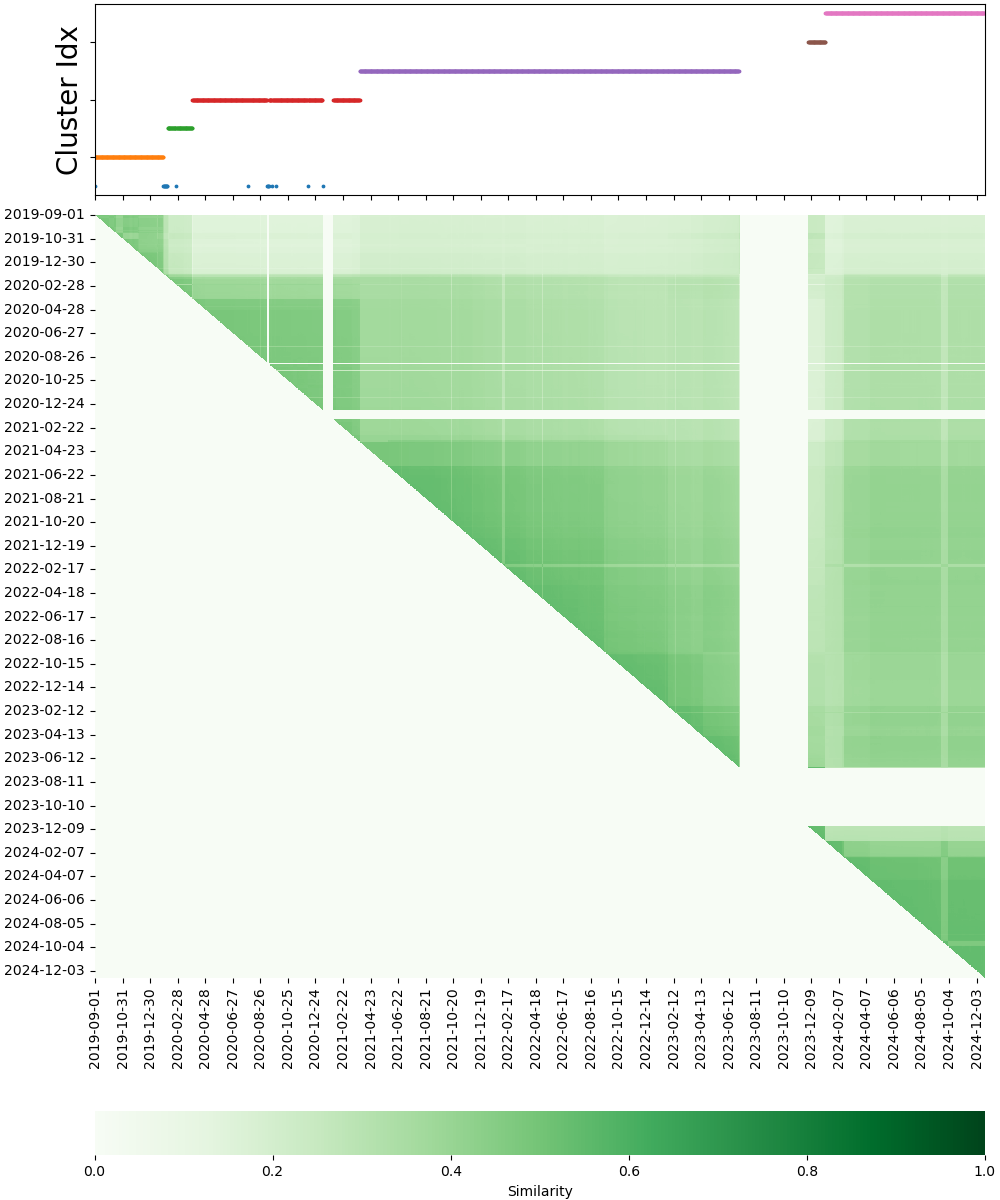}
      \draw[coordinate label = {(i) at (0.12,0.85)}];s
      \draw[coordinate label = {(ii) at (0.15,0.91)}];
      \draw[coordinate label = {(iii) at (0.25,0.94)}];
      \draw[coordinate label = {(iv) at (0.52,0.96)}];
      \draw[coordinate label = {(v) at (0.82,0.94)}];
      \draw[coordinate label = {(vi) at (0.9,0.96)}];
	    \draw[thick,blue] (0.52,0.3) rectangle (0.82,0.535) 
	        node[below left,white,fill=blue]{\small 1};

      \path (0.6,0.4)node(x){}(0.6,0.5)node(y){};
      \draw[thick,blue] (x) -- (y);

      \path (0.67,0.37)node(x){}(0.67,0.48)node(y){};
      \draw[thick,blue] (x) -- (y);

      \path (0.7,0.34)node(x){}(0.7,0.46)node(y){};
      \draw[thick,blue] (x) -- (y);

      \draw[coordinate label = {iv-a at (0.554,0.46)}];
      \draw[coordinate label = {iv-b at (0.63,0.4)}];
      \draw[coordinate label = {iv-c at (0.66,0.37)}];
      \draw[coordinate label = {iv-d at (0.74,0.36)}];
    \end{annotationimage}
\end{tabular}
}

\caption{\broot catchments from 2019-09 to 2024-12.  Dataset: \broot/Verfploeter. Zoom in the rectangle 1 in \autoref{fig:broot_zoomed_in_2023} to show latency changes. }
	\label{fig:broot_2019_2025}

\end{figure}
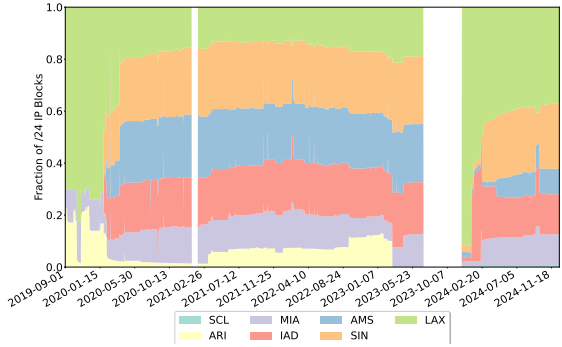
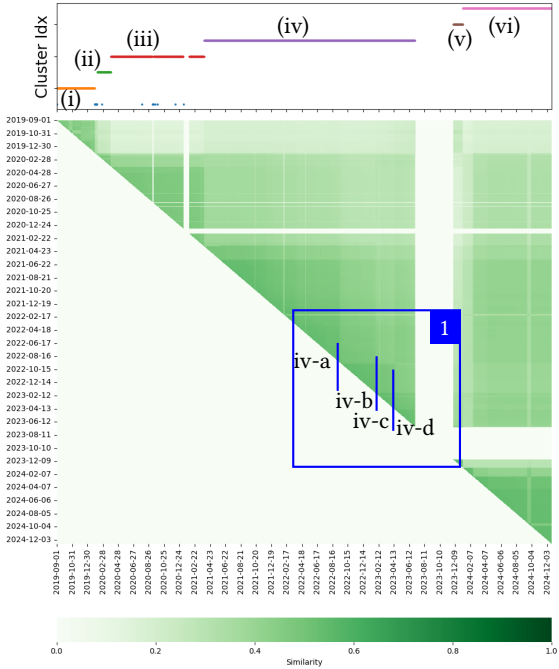

DNS services often use IP Anycast to provide service
  from many locations.
Anycast both reduces latency to clients by associating each with a nearby server,
  and helps protect during DDoS events~\cite{Moura16b}.
Latency directly affects clients
  and is often used as a metric to compare service providers,
  so DNS operators monitor latency and routing~\cite{Bellis15a,Schmidt17a},
  since inefficient routing can result in high latency
  due to anycast polarization~\cite{Moura22a}.

We next examine five years of \broot DNS anycast data
  to demonstrate how our vectors help identify trends and changes,
    collected with Verfploeter (\autoref{sec:data_collection_anycast}).

\subsubsection{Changes in Routing Modes}
	\label{sec:modes_broot}

\autoref{fig:verfploeter_broot_heatmap_20190901.20250501} 
	shows a heatmap of vectors similarity over five years.
In this graph, both axes show dates of observations,
  and each coordinate at $(t, t')$ compares the similarity
  of vectors between those dates.
We automatically discover six common routing modes (mode (i) to mode (vi))
  and these are visible as several darker triangles.
(The blank region from 2023-07-05 to 2023-12-01 is a collection outage.)

To understand \emph{why} these clusters appear similar,
  \autoref{fig:verfploeter_broot_stackplot_20190901.20250501}
  shows a stack-plot of the number of networks in each catchment.

First, 
 overall routing is relatively stable,
 sometimes for years at a time,
 with minor varations inside each cluster.

Routing is stable for the first five months, 
	from 2019-09 to 2020-02 (mode (i), with the similarity $\Phi$ in [0.24, 0.49]).

There is a new mode (ii) from 2020-02 to 2020-04,
	with $\Phi$ in [0.26, 0.49], and $\PhiModes{i}{ii} = [0.11, 0.48]$.
This change corresponds to adding three new anycast sites
  \SINstack, \IADstack, and \AMSstack to \broot.
We then see a different mode (iii), from 2020-04 to 2021-03,
	with $\Phi$ in [0.2, 0.54], and $\PhiModes{ii}{iii} = [0.18, 0.48]$.
This is becuase around 70\% clients used to go \LAXstack were routed to \AMSstack, \IADstack and \SINstack.
The longest-lasting mode (iv) covers 2021-03 to 2023-07, with $\Phi$ in [0.24, 0.49].
On 2023-07, routing changes from mode (iv) to mode (v),
	$\PhiModes{iv}{v} = [0.18, 0.54]$.
This corresponds with the move of \ARIstack to a new location in the same country.
Interestingly, mode (v) is somewhat like 
	the original routing mode (i) in 2019-09,
        with $\PhiModes{i}{v} = 0.31$,
        more so than its immediate neighbors
        ($\PhiModes{iv}{v} = 0.22$, and 
        $\PhiModes{v}{vi} = 0.17$).
This similarity is because        
	\LAXstack serves most clients in both (i) and (v).
This ability to quantify similarity and discover
  connections such as between (i) and (v) is a Fenrir design goal.

The opportunity that routing modes provide to operators is to see
  trends over time, and to identify similarities in routing and differences.
Similarities can give operators confidence in what latency their users will experience,
  and changes in similarity that are not accompanied by known
  changes by the operators suggest that 
  some changes in third-party routing have affected their traffic.
This global picture helps guide daily performance monitoring,
  providing routine that shows the ``root causes'' behind measurements of customer latency.

\begin{figure}
  \begin{tabular}{l}
    \imagelabelset{coordinate label back = none, coordinate label text= black, coordinate label font = \small}
    \begin{annotationimage}{width=1\linewidth,clip}{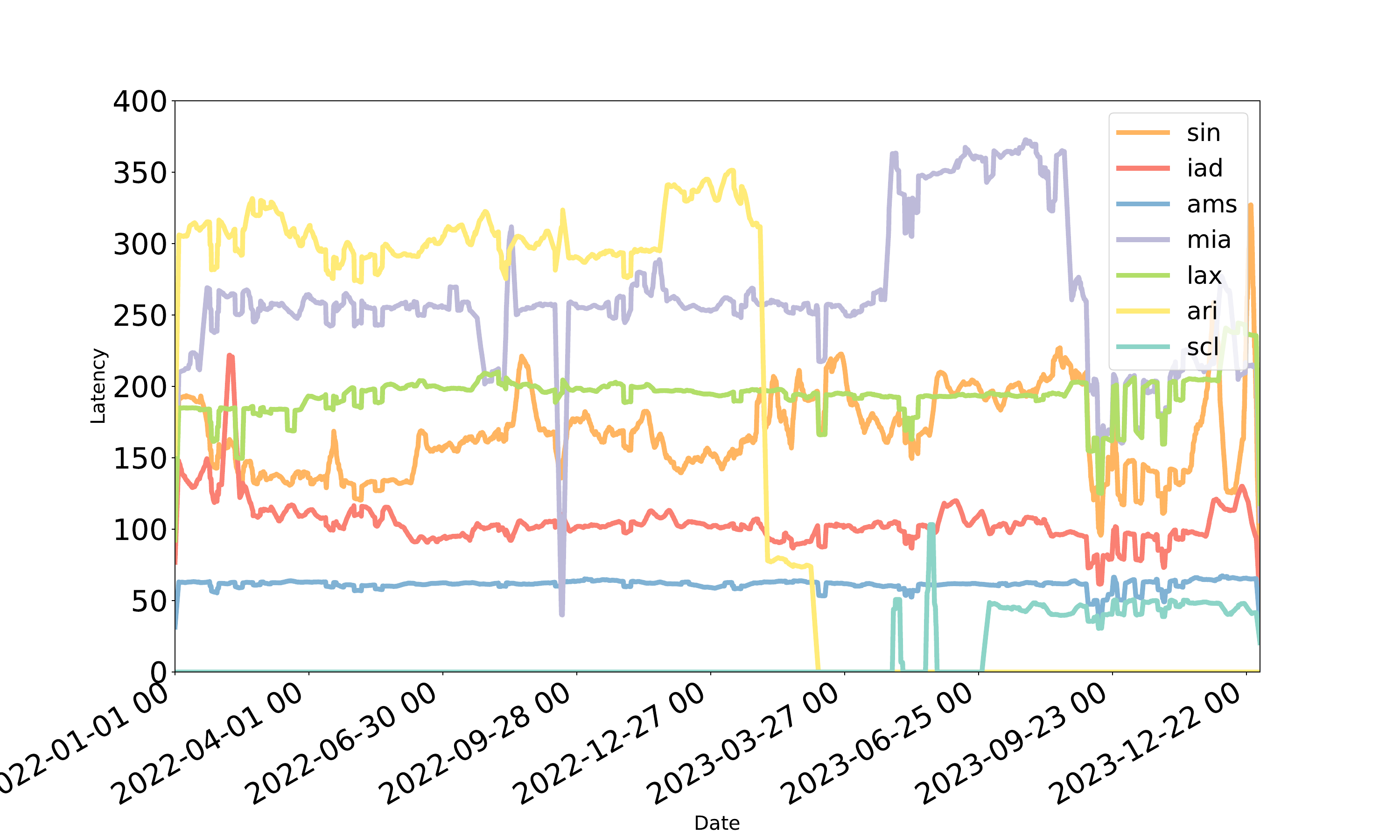}
    \end{annotationimage}
\end{tabular}
\caption{The 90th percentile of latency of \broot, from 2022-01-01 to 2023-12-31. Dataset: RIPE-Atlas} 
\label{fig:broot_zoomed_in_2023}
\end{figure}

\subsubsection{Changes in Latency}
    \label{sec:latency_broot}

While operaters influence routing and catchments (\autoref{sec:modes_broot}),
  \emph{latency to customers} is what really matters.
We next show how similarities and changes in routing modes correspond to potential
  changes in latency.

As a case study, we examine \broot from 2022-01 to 2023-12,
  the blue-boxed region [1] in \autoref{fig:broot_2019_2025},
  which shows catchment changes for all networks as measured with Fenrir/\broot/Verfploeter.
Mode detection shows two modes over this period, (iv),
  followed by an unfortunate gap observation from August to October,
  then a new mode (v).

Close observation of the figure also shows several routing changes
  that are smaller than our threshold to declare a new mode.
We mark these as (iv.a) to (iv.d) on \autoref{fig:broot_2019_2025}.
We identify (iv.c) ending on 2022-09-16,
	(iv.b) ending on 2023-02-12,
  (iv.c) ending on 2023-04-13,
  (iv.d) ending on 2023-07-05.
These events show up as small but significant changes in $\Phi$:
  each mode has an internal $\Phi$ around [0.475, 0.54],
  but with $\PhiModes{iv.a}{iv.b} = [0.33, 0.54]$,
	$\PhiModes{iv.b}{iv.c} = [0.39, 0.54]$, 
  and $\PhiModes{iv.c}{iv.d} = [0.4, 0.54]$.   


To understand what these changes represent, we looked at public announcements
about \broot service
\ifisanon
changes~\cite{BRoot23bAnon},
\else
changes~\cite{BRoot23b},
\fi
that service in South America was changing, with site \ARIstack shutting down on 2023-03-06
and \SCLstack starting in 2023-05.

We expect operators to watch Finrir,
notice changes like those we note above,
and for changes that are large enough, check on latency measurements.
We therefore gather latency data for \broot for each catchment net
from RIPE Atlas to stand in for an operator's latency observations.

\autoref{fig:broot_zoomed_in_2023} shows p90 latency for each of \broot's catchments.
We can see the service changes result in changes to catchments and latency:
first the \ARIstack catchment vanishes on 2023-03-06.
\ARIstack provided latency over 200\,ms due to a few North American and European networks
being routed to it,
but its latency drops to zero when it is shut down.
This
After two months, \SCLstack appears, but only briefly on 2023-05-01 and 2023-05-24.
Our understanding from the discussion with operators is that these were
temporary routing experiments as they were optimizing routing.
Finally, on 2023-06-29, \SCLstack resumed operation,
providing very low latency.

This example shows how changes can be detected in Finrir
via changes in $\Phi$ shown in \autoref{fig:broot_2019_2025},
and the operator uses them to trigger latency investigation
(as we show in \autoref{fig:broot_zoomed_in_2023}).
Although, in this case, the operator was aware of changes,
if the changes had happened networks further upstream,
Fenrir's ability to detect changes would be critical.

\subsection{How Quickly Do Top Website's Front-ends Shift?}

We run Fenrir on two top websites: Google's front page and Wikipedia's home page.
Our analysis shows Google employs aggressive deployment~\cite{Govindan16a} 
	to frequently change routing;
	on the other hand,
Wikipedia's clients tend to go to the same site.

\subsubsection{Google}
\label{sec:results_google}

Google is certainly a top-ten website,
so we study them as an example of how such websites behave.
Google began two policies that have been widely emulated:
first, they aggressively deployed thousands of front ends around
the Internet~\cite{Calder13a}, so they could place their content near end-users,
often directly peering with ``eyeball'' networks
and enterprises (as we saw in \autoref{sec:results_enterprise}).
Second, Google frequently updates production services,
and migrates customers between front ends, so this churn
is not generally visible~\cite{Govindan16a}.

We study how Google presents itself to customers
by looking at their front web page
and using the EDNS/Client-Subnet extension to see
what server is assigned to global networks.
These servers are the Google customer catchments.

\begin{figure}
  \includegraphics[width=1\linewidth,clip]{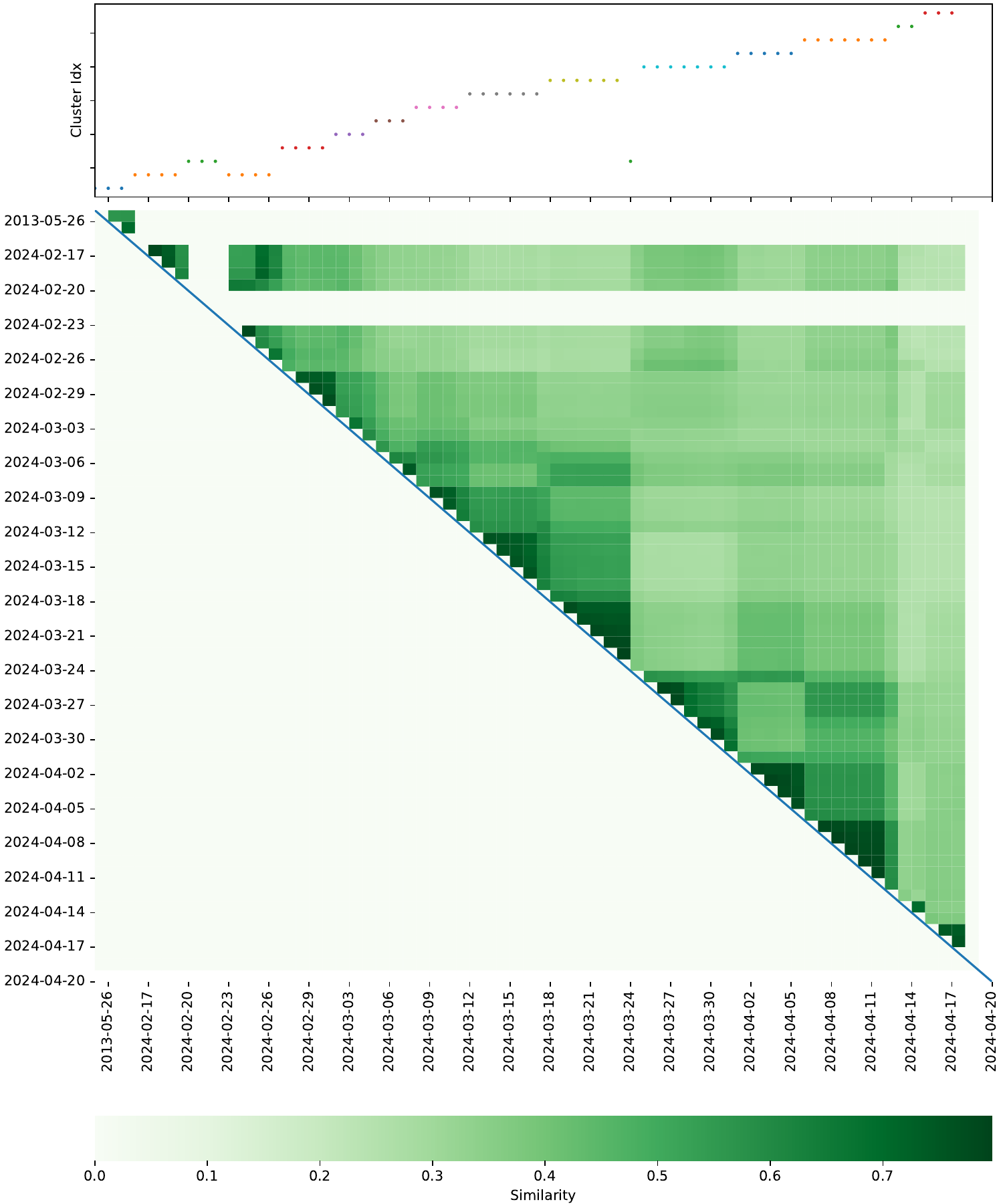}
  \caption{Heatmap of routing changes of Google, from 2013-05-26 to 2013-05-28 and from 2024-02-17 to 2024-04-21. Dataset: Google/EDNS-CS}
  \label{fig:EDNS.20240217.20240421.heatmap}
\end{figure}

\paragraph{Change in Routing Modes:}
\autoref{fig:EDNS.20240217.20240421.heatmap}
shows the heatmap of routing changes of Google's front-ends.
We have data for two discontiguous periods,
three days starting 2013-05-26~\cite{Calder13a}, 
and then 60 days starting 2024-02-21. 
The heatmap shows the full range of routing modes similarity,
and the top three lines are eight years before the remainder of the graph.

The overall trend shows a fairly strong mode,
with strong similarities lasting about seven days
and significant differences with the next period.
$\Phi$ within each week are pretty high, around 0.79,
while between periods $\Phi$ are only around 0.25.
That mode suggests regularly scheduled changes
corresponding with the work week.
Some sequences of periods show larger similarities,
such as the periods starting 2024-03-03, and then -11 and -18,
and the three periods starting 2024-03-24, -04-01, and -04-07.

As one would expect,
data from 2013 is quite different from today.
The three days starting 2013-05-26 have no similarity
with any of the modern infrastructure---Google
has completely changed its front-end infrastructures after ten years.

Overall, these results show Google's policy of aggressive deployment~\cite{Govindan16a}
means clients often change the front ends they are using.

\subsubsection{Wikipeida}
	\label{sec:results_wikipedia}

Wikipedia
  is another top-10 website.
Wikipedia is a non-profit
  and operates seven Front-end sites, each serving clients around the world,
  with clients associated with site based on client location.
With a much smaller footprint and different site-slection method,
  Wikpedia complements our examination of Google.

\begin{figure}

\subfloat[The aggregated catchment distribution of Wikipedia.]{
	\label{fig:edns_wiki_stackplot}
  \begin{tabular}{l}
    \includegraphics[width=1\linewidth,clip]{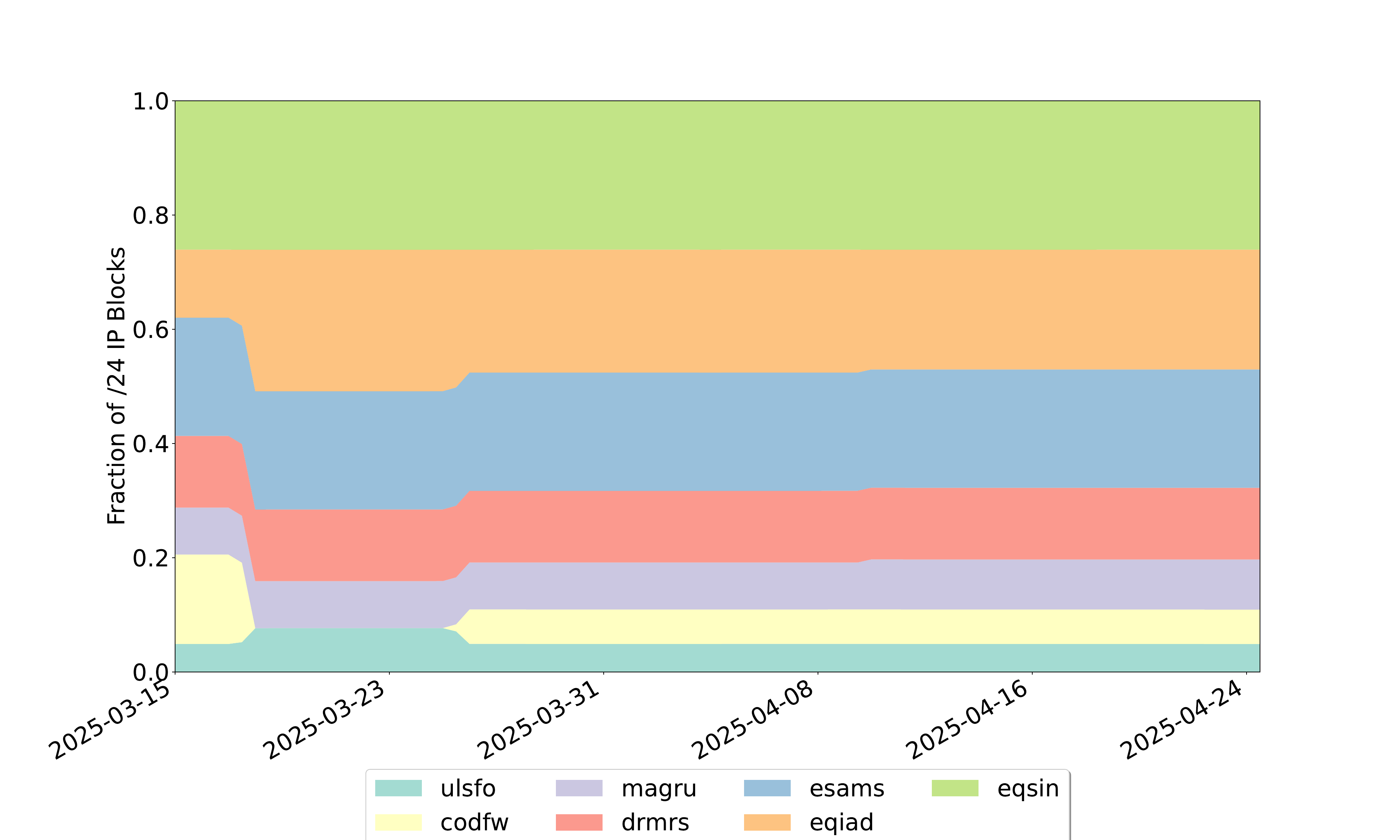}
\end{tabular}
}

\subfloat[Heatmap comparing vectors of Wikipedia.]{
 \label{fig:edns_wiki_heatmap}
  \begin{tabular}{l}

    \imagelabelset{coordinate label back = none, coordinate label text= black, coordinate label font = \small}
    \begin{annotationimage}{width=0.9\linewidth,clip}{FIG/Wikimedia.heatmap.20250315.20250330.pdf}
      \draw[coordinate label = {(i) at (0.13,0.86)}];
      \draw[coordinate label = {(ii) at (0.22,0.91)}];
      \draw[coordinate label = {(iii) at (0.9,0.96)}];
    \end{annotationimage}

	\end{tabular}
	}

	\caption{Wikipedia catchments from 2025-03-15 to 2025-04-26. Dataset: Wiki/EDNS-CS. }
	\label{fig:wiki_plot}
\end{figure}

We measure Wikipedia following \autoref{sec:data_collection_top_website}
  and show Wikipedia catchments in \autoref{fig:wiki_plot}.

\paragraph{Change in Routing Modes:}
We see that Wikipedia's
  catchments are very stable over this observation period,
  with each mode showing $\Phi$ in [0.93, 0.95].

The exception is mode (ii),
  the week starting 2025-03-19.
Fenrir shows $\PhiModes{i}{ii} = [0.79. 0.94]$,
  so about 20\% of networks shift when one site goes offline.
\autoref{fig:edns_wiki_stackplot} shows the absence of \colorbox{lightyellow}{codfs},
  with 75\% of its clients going to \colorbox{lightorange}{eqiad}
  and 25\% to \colorbox{seablue}{ulsfo}.

We also see that when \colorbox{lightyellow}{codfw} returns on 2025-03-26,
  the new mode is slightly different from before:
   $\PhiModes{i}{iii} = [0.8. 0.94]$,
   since only 30\% of  \colorbox{lightyellow}{codfw}'s original client return.

The Wikipedia operators knew about \colorbox{lightyellow}{codfw} going offline.
This event is confirmed on Wikipedia's public dashboard~\cite{Wikimedia25a},
	indicating \colorbox{lightyellow}{codfw} was drained on 2025-03-19,
	and was back online on 2025-03-26.
Fenrir helps quantify how different the post-event catchments
  are compared to pre-event.
For websites that use anycast instead of geographic catchment selection,
  Fenrir also helps discover catchment changes that result from third-party routing changes.

\section{Related Work}

Our work builds on prior work studying routing and anycast
to optimize routing.

\textbf{Routing optimization and measurement for enterprises and hypergiants:}
A number of commercial tools optimize routing for enterprises~\cite{Rexford06a}
	and to provide general reporting~\cite{ThousandEyes21a}.
Hypergiants optimize
	their internal WANs~\cite{Jain13a}
	and their IXP- and customer-facing links~\cite{Ager12a,Jain13a,Schlinker17a,Yap17a}.
Often this prior work depends on global knowledge of ISP or hypergiant traffic
	and detailed access to the control plan (routing) and data plane (traffic)
	information.
We share these goals,
	but focus on observing changes
	with unprivileged tools (DNS and traceroutes).

A great deal of routing analysis considers control-plane measurements
	from BGP peering sessions.
Very close to our work is work from Boston University
	uses hierarchical clustering and similarity heatmaps
	to study routing to Google, an anycast DNS, and an IXP~\cite{Comarela16a}.
Unlike our work, their goal is to identify ``unusual routes'' and hijacks in
	control-plane data,
We currently use data-plane sources to detect routing stability in several ways.
Others study BGP hijacking with RouteViews (control-plane) data~\cite{Shah16a},
	to protect applications~\cite{Sun21a},
	or to evaluate risks to specific applications like Tor~\cite{Sun20a}.
While in principle, our approach could use control-plane information as
	a data source, demonstrating that it is future work.
We share the goal of these control-plane tools to understand routing,
	but they often go deeper into specific threats (hijacking)
	or applications (Tor).

RouteViews~\cite{RouteViews00a} and RIPE RIS~\cite{Ripe11a}
  are a long-term BGP data repository,
  and
  several general control-plane analysis tools have been deployed~\cite{Chi08a,Orsini16a}
  using it or similar information.
These tools have been applied to
  track Internet growth~\cite{Oliveira07a}
  and structure~\cite{Oliveira08a}.
We instead use data-plane information to evaluate how routing changes
  affect specific services.

\textbf{Anycast measurement and optimization:}
IP anycast~\cite{Partridge93a}
  is widely used today by DNS~\cite{Hardie02a,Colitti06a} and CDNs~\cite{Schomp20a,Koch21a}
  to minimize latency~\cite{Schmidt17a} and manage DDoS~\cite{Moura16b,Rizvi22a}.
Users of IP anycast have developed a special 
  tools to evaluate destinations with active probing~\cite{Woolf07a,Vries17b,Sommese20a}
  and passive analysis~\cite{Bian19a,Moura22a}.
They have applied these tools to 
  identify and resolve specific problems like anycast polarization~\cite{Moura22a,Rizvi24a},
  suboptimal routing~\cite{Bellis15a,Wei20a}
  and to evaluate specific applications like regional anycast systems~\cite{Zhou23a}
  and enterprise ingress traffic~\cite{Koch23a}.
These approaches identify specific performance questions
  and optimize anycast networks in different ways.
Our work instead focuses broadly on anycast situational awareness,
  and could serve as a trigger to prompt using the above approaches
  to investigate current user performance or opportunities to optimize anycast.

\textbf{Measuring web services:}
There is a rich field in measuring web services,
  with both commercial approaches (for example,~\cite{ThousandEyes21a})
  and research approaches.
Our work instead focuses on routing, not web performance specifically.
However, we benefit from the EDNS/client-subnet methods
  begun to study Google~\cite{Calder13a}.

\section{Conclusion}

This paper presented present Fenrir,
a tool to evaluate recurring routing results,
with applications to anycast services,
multi-homed networks,
and top websites deployed across many front-ends.
We describe how Fenrir can ingest data
from active probing taken from several different measurement tools
such as RIPE Atlas, traceroutes, and website lookups.
After data cleaning, Fenrir reduces this data to a \emph{routing vector}
that documents how user networks reach the service of interest,
defining \emph{catchments} that can optionally be weighted to reflect
service performance.
We then describe three techniques for routing vector analysis:
pairwise comparison, hierarchical clustering, and visualization
through all-pairs heatmaps and vector transition matrices.

We apply Fenrir to evaluate routing in three specific applications:
\broot as an example of a service using IP anycast,
routing out of a multi-homed enterprise, and google as an example website served by an extensive network of front-ends.
We leverage existing datasets spanning five years,
and collect customized new monthly datasets to study these systems.
We validate our approach for \broot anycast,
using operator-provided ground truth,
showing perfect recall (1) and good accuracy (0.84).
Precision is lower (0.7), but because Fenrir's goal
is to detect not only operator-known events,
but also to detect \emph{third-party} changes to routing
of which operators would otherwise be unaware.

Applying our tool to these applications,
clustering identifies several common ``modes'' in
\broot anycast, discovering that about one-third of catchments between 2019 and 2024 match.
Based on eight months of data of the multi-homed enterprise \OurUniversity,
we show that its routing is very stable at its immediate upstreams;
however, a huge routing change results in at most 90\% of catchment changes.
Finally, examination of Google's homepage as an example of
top website captures its aggressive policy for rapid service evolution,
particularly compared to our other subjects.
Users see very similar front-ends (79\% similar) for a week,
but only 25\% similar across weeks.
On the other hand, Wikipedia's routing is relatively stable,
except the time when one site was down;
after the site was up again,
the new routing result is only 80\% similar to the previous one.
These several applications show that Fenrir is a useful general tool
to understand Internet-wide routing.    

\ifisanon
\else
\vspace{2ex}
\noindent \textbf{Acknowledgments:}
This work was partially supported by NSF through
  grants CNS-2212480 
  and CNS-2319409. 

\fi

\label{page:last_body}

 \clearpage

\bibliographystyle{ACM-Reference-Format}
\bibliography{paper}

\newpage

\appendix

\section{Changes in Upstreams Providers of the Enterprise Network}

  \begin{figure*}
  	\centering
  	\begin{minipage}[b]{.45\linewidth}
    \adjincludegraphics[width=1\linewidth,trim={0 0 0 100},clip]{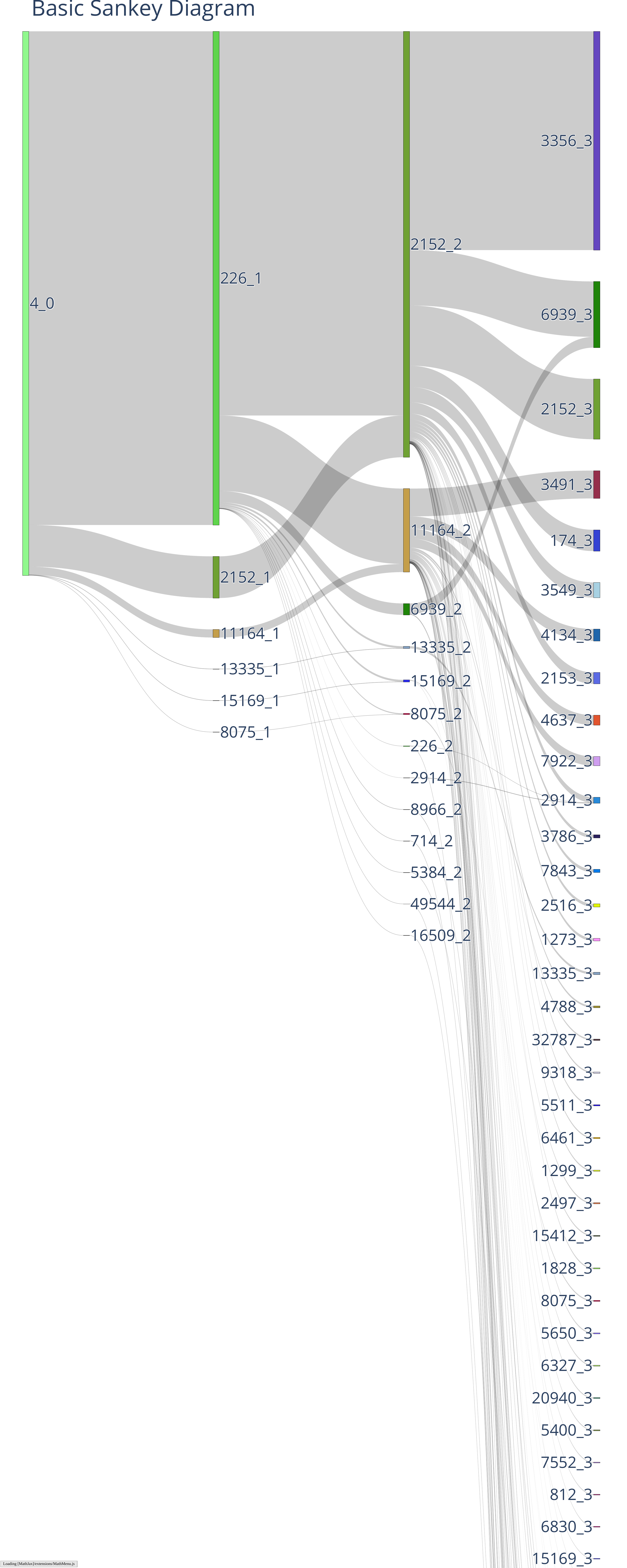}
    \caption{The flow topology of a multi-homed enterprises before the routing change, 2025-01-14.}
    \label{fig:sankey_enterprise_before}
  	\end{minipage}
  	\hfil
  	\begin{minipage}[b]{.45\linewidth}
      \adjincludegraphics[width=1\linewidth,trim={0 0 0 100},clip]{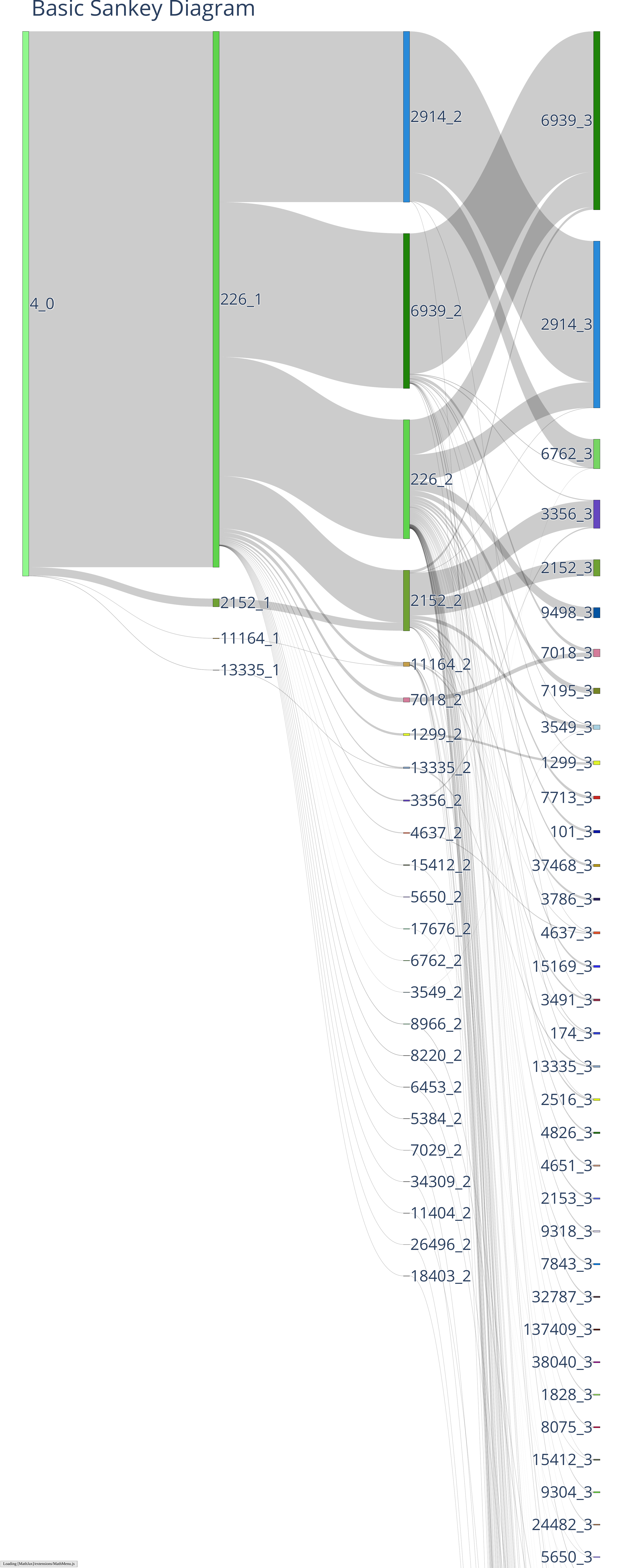}
      \caption{The flow topology of a multi-homed enterprises after the routing change, 2025-01-20.}
      \label{fig:sankey_enterprise_after}
  	\end{minipage}%
\end{figure*}

We use the Sankey diagram to study how the flow has changed starting from the enterprise network,
\autoref{fig:sankey_enterprise_before} and \autoref{fig:sankey_enterprise_after} illustrate
the number of networks served by each upstream provider at hop 1-4.
The number labeled at each node is the AS number of the upstream provider.

At hop 2 (as labeled \_1 in the graph),
on 2025-01-14, 8\% destination networks were routed by AS 2152;
on 2025-01-20, the percentage dropped to 1.5\% after the huge routing change.
At hop 3 (as labeled \_2 in the graph),
on 2025-01-14, 80\% destination networks were routed by AS 2152;
on 2025-01-20, the percentage dropped to 13\% after the huge routing change.
Also, those changed networks were instead routed by AS 2914 (31\%), AS 6939 (29\%), and AS 226 (22\%).
At hop 4 (as labeled \_1 in the graph),
the change in routing result is more significant,
as it is further from the enterprise.

By using Sankey diagram to visualize the routing result,
we can determine what upstream providers are influenced when routing changes.

\label{page:last_page}

\end{document}